\documentclass[12pt,preprint] {aastex}
\DeclareMathAlphabet{\mathsc}{OT1}{cmr}{m}{sc}
\def\testbx{bx}%
\DeclareRobustCommand{\ion}[2]{%
\relax\ifmmode
\ifx\testbx\f@series
{\mathbf{#1\,\mathsc{#2}}}\else
{\mathrm{#1\,\mathsc{#2}}}\fi
\else\textup{#1\,{\mdseries\textsc{#2}}}%
\fi}

\begin{document}
 
\title {The Supernova Rate-Velocity dispersion Relation in the Interstellar Medium}

\altaffiltext {1}{Max-Planck-Institut f\"{u}r Astronomie, K\"onigstuhl 17, D-69117 Heidelberg, Germany; dib@mpia.de, bell@mpia.de}
\altaffiltext {2}{University Observatory Munich, Scheinerstrasse 1, D-81679 Munich, Germany; burkert@usm.uni-muenchen.de }
\altaffiltext {3}{Centro de Astronom\'{i}a y Astrof\'{i}sica, UNAM, Apdo. 72-3 (Xangari), 58089 Morelia, Michoac\'{a}n, Mexico}
\author{Sami Dib\altaffilmark{1,2,3}, Eric Bell\altaffilmark{1} and Andreas Burkert\altaffilmark{2}}

\begin{abstract} 
 We investigate with three-dimensional numerical simulations of supernova driven turbulence in the interstellar medium (ISM) the relationship between the velocity dispersion of the gas and the supernova rate and feedback efficiency. Our simulations aim to explore the constancy of the velocity dispersion profiles in the outer parts of galactic disk at $\sim 6-8$ km s$^{-1}$, and the transition to the starburst regime i.e., high star formation rates associated with high velocity dispersions. With our fiducial value of the supernova feedback efficiency (i.e., $\epsilon=0.25$ corresponding to an injected energy per supernova of $0.25 \times 10^{51}$ ergs), our results show that (a) supernova driving leads to constant velocity dispersions of $\sigma \sim 6$ km s$^{-1}$ for the total gas and $\sigma_{\ion{H}{i}} \sim 3$ km s$^{-1}$ for the \ion{H}{i} gas, independent of the supernova rate, for values of the rate between 0.01 and 0.5 the Galactic value ($\eta_{G}$), (b) the position of the transition to the starburst regime (i.e., location of sharp increase in the velocity dispersion) at around SFR/Area $\simeq 5 \times 10^{-3}-10^{-2}$ M$_{\odot}$ yr$^{-1}$ kpc$^{-2}$ observed in the simulations, is in good agreement with the transition to the starburst regime in the observations (e.g., NGC 628 and NGC 6949), (c) for the high SN rates, no \ion{H}{i} gas is present in the simulations box, however, for the total gas velocity dispersion, there is good agreement between the models and the observations, (d) at the intermediate SN rates ($\eta/\eta_{G} \sim 0.5-1$), taking into account the thermal broadening of the \ion{H}{i} line helps reach a good agreement in that regime between the models and the observations, (e) for $\eta/\eta_{G} < 0.5$, $\sigma$ and $\sigma_{\ion{H}{i}}$ fall below the observed values by a factor of $\sim 2$. However, a set of simulation with different values of $\epsilon$ indicates that for larger values of the supernova feedback efficiencies, velocity dispersions of the \ion{H}{i} gas of the order of $5-6$ km s$^{-1}$ can be obtained, in closer agreement with the observations. The fact that for $\eta/\eta_{G} < 0.5$, the \ion{H}{i} gas velocity dispersions are a factor $\sim 2$ smaller than the observed values could result from the fact that we might have underestimated the supernova feedback efficiency. On the other hand, it might also be an indication that other physical processes couple to the stellar feedback in order to produce the observed level of turbulence in galactic disks.  
     
\end{abstract} 

\keywords{Galaxies --- Galaxies : velocity dispersion--ISM : supernova--turbulence--instabilities: thermal instability}

\section{INTRODUCTION}\label{intro}

It is now a well established fact that the interstellar medium (ISM) in galactic disks is turbulent (Larson 1981; Scalo 1987; Dickey \& Lockman 1990; Elmegreen \& Scalo 2004). In most spiral galaxies, and after correcting for instrumental effects, the vertical velocity dispersion derived essentially from \ion{H}{i} observations is observed to vary radially from $\sim 12-15$ km s$^{-1}$ in the central parts to $\sim 4-6$ km s$^{-1}$ in the outer parts (van de Kruit \& Shostak 1982 for NGC 3938; Shostak \& van der Kruit 1984 for NGC 628; Dickey, Hanson \& Helou 1990 for NGC 1058; Kamphuis \& Sancisi 1993 for NGC 6964; Rownd, Dickey \& Helou 1994, for NGC 5474; Meurer, Mackie \& Carignan 1994 and Meurer et al. 1996 for NGC 2915; de Blok \& Walter 2005 for NGC 6822). In most cases, the latter values exceed those expected by the thermal broadening of the \ion{H}{i} emission line. As an example, the radial dependence of the velocity dispersion in NGC 1058 is shown in Fig.~\ref{fig1}. Fig.~\ref{fig2} shows the behavior of the characteristic velocity dispersion of a sample of galaxies when plotted versus their characteristic star formation rate. The galaxies shown in Fig.~\ref{fig2} correspond to a sample where both the velocity dispersion and the star formation rate were available in the literature. Similarly to the observed velocity dispersion radial profiles, the data plotted in Fig.~\ref{fig2} argues for the existence of a minimum level of turbulence in galactic disks. The case of NGC 2915 is particularly intriguing as it has an extended \ion{H}{i} disk with a constant velocity dispersion, even in the outer parts, where stars do not form (Meurer et al. 1996). 

Many physical processes might contribute to the driving of the turbulence in the ISM, acting on different scales and injecting different amounts of kinetic energy into the medium. These driving mechanism can be related to stellar activity (e.g., ionizing radiation from \ion{H}{ii} regions, jets from young stellar objects, stellar winds, supernova explosions), or to gas hydrodynamical or magnetohydrodynamical instabilities (e.g., thermal and gravitational instabilities, magnetorotational instability, Kelvin-Helmotz, Rayleigh-Taylor and Parker instabilities) or, eventually, to the complex interaction between these processes. Mac Low \& Klessen (2004) showed, using simple analytical estimates, that the global energy input into the ISM of the Galaxy from supernova explosions (SNe) is much larger than from any of the above mentioned processes. However, their estimates are global ones which do not compare the efficiency of each process at different galactic radii and the possible interactions between two or several processes. Other, yet poorly explored, sources of kinetic energy injection into the ISM of some galaxies might reside in perturbations occurring on galactic scales (e.g., frequent minor mergers, tidal interactions with satellite galaxies and ram pressure effects). 

In the past decade, some of the above mentioned energy injection mechanisms into the ISM have been investigated by means of numerical simulations. Wada \& Norman (1999,2001), Wada, Spaans \& Kim (2000) and Wada, Meurer \& Norman (2002) investigated the evolution of thermal and gravitational instabilities (TI and GI, respectively) in galactic disks. Wada, Meurer \& Norman (2002) showed, using high resolution two-dimensional simulations of an NGC 2915 like disk that the turbulent energy spectra can be maintained in a quasi-stationary state in the absence of stellar feedback. In the latter simulations, the kinetic energy decay and radiative cooling are compensated by the interaction of the galactic shear with the gas self-gravity. Another indication that energy might be injected into the ISM of low star forming galaxies on large scales before cascading to smaller scales via thermal and gravitational instabilities is provided by the analysis of their \ion{H}{i} gas morphology. By comparing the \ion{H}{i} emission map of Holmberg II (Ho II) to synthetic \ion{H}{i} maps of driven turbulence which include also cooling, heating and gravity, Dib \& Burkert (2005) found that kinetic energy is injected into the ISM of Ho II on a scale of $\sim 6$ kpc, which is much larger than the scale implied by SN driving. Galactic shear might play a role in supporting turbulence in Ho II, however, the comet shape structure of the \ion{H}{i} gas in Ho II (Bureau et al. 2004) suggests that turbulence might be also partially induced by the effect of tidal interactions with Ho II's two satellite galaxies. 

The coupling of the galactic shear to magnetic fields can trigger the magnetorotational instability (MRI) (Balbus \& Hawley 1991). MRI has been the scenario invoked by Selwood \& Balbus (1999) to explain the constancy of the velocity dispersion in the outer parts of NGC 1058. The main argument of Selwood \& Balbus (1999) against a SN-driven ISM in the outer regions of NGC 1058, in addition to the fact that column densities are low and star formation is inefficient, is the observed uniformity of the \ion{H}{i} velocity dispersions (Dickey, Hanson \& Helou 1990) whereas \ion{H}{ii} regions in NGC 1058 are primarily observed in narrow spiral arms (Ferguson et al. 1998). Local, three-dimensional simulations with a single phase medium by Kim, Ostriker \& Stone (2003) of the MRI result in velocity dispersions of the order $\sim 1.6-3.2$ km s$^{-1}$. Global one-phase medium simulations by Dziourkevitch et al. (2004) and Dziourkevitch (2005) show that some amount of turbulent motions can be created by the MRI in galactic disks. However, this turbulence is mostly located in the inner parts of the disk (the inner 4 kpc) and the velocity dispersion does not exceed in that case $\sim 3.5-4$ km s$^{-1}$, dropping quickly to very small values at larger radii. Two and three-dimensional MRI simulations in a medium affected by the TI but with no stellar feedback have been presented by Piontek \& Ostriker (2004,2005). In the three-dimensional models, the latter authors show that the velocity dispersion depends on the average density of the medium (larger dispersions at lower densities). At the lowest density  they have considered (0.25 cm$^{-3}$), the one-dimensional velocity dispersion is found to be 2.2, 2.2 and 1.1 km s$^{-1}$ for the three directions of the box, respectively ($\sim$ 4.5 km s$^{-1}$ for the three components if thermal broadening is not subtracted). The latter values are a factor of 3-6 lower than the observational values ($\sim $ 1.5-2 if thermal broadening is not subtracted). Recent simulations of the MRI in a multiphase medium with star formation included show that the MRI might be completely suppressed/overwhelmed by stellar feedback (M. Korpi, private communication).

 Another major source of energy input into the ISM is the stellar energy feedback, particularly from massive stars. The latter can be delivered to the ISM in the form of ionizing radiation and stellar winds from O and B stars (Kessel-Deynet \& Burkert 2003) and from clustered or field supernova explosions (McKee \& Ostriker 1977). Two and three-dimensional numerical models of SN explosions models in the ISM have been presented in the literature (Rosen \& Bregman 1995; Korpi et al. 1999a,1999b; Gazol-Pati\~{n}o \& Passot 1999; de Avillez 2000; de Avillez \& Berry 2001; Avila-Reese \& V\'{a}zquez-Semadeni 2001; Kim et al. 2001; Kim 2004; de Avillez \& Breitschwerdt 2004,2005; Slyz et al. 2005; Mac Low et al. 2005). In these models, the authors have focused on problems like the evolution of SNe bubbles and their outburst through the galactic disk, the halo-disc interaction, the vertical scale heights and volume filling factors of the different gaseous phases and the effects of supernova explosions on the Galactic dynamo (Ferri\`{e}re 1992a,1992b,1998a,1998b,2000). However, the existence of a correlation between the SN rate and the velocity dispersion of the gas has not been investigated so far. A different interpretation of the velocity dispersion in the outer regions of galactic disks is discussed by Schaye (2004), which argues that if the surface density in the outer parts of the disk falls below a critical value (i.e., column density $Log(N_{H}) < 27.5$ cm$^{-2}$), the medium can be efficiently heated to a temperature of $T \sim 8000$ K by a background of ultraviolet (UV) ionizing radiation, both of galactic and extragalactic origin. Thus, thermal broadening could account for the observed level of line broadening in the outer parts of galactic disks, leaving little room for a dynamical, turbulent component. This hypothesis can be appropriately tested if the level of UV radiation, particularly the extragalactic component, and the temperature radial profiles in galaxies would be better constrained from observations. 

 The aim of this paper is to assess how much of the velocity dispersion observed in the ISM of galaxies can be due to SN feedback for various values of the SN rate and feedback efficiency. A particular point of interest is to understand the constancy of the velocity dispersion in galactic disks at different radii where the star formation rate is expected to decrease with increasing radius from the galactic center as predicted by many empirical star formation laws (Schmidt 1959,1963; Kennicutt 1998a,1998b; Dopita \& Ryder 1994; Prantzos \& Silk 1998; see also Li et al. 2005a,2005b who finds, as an explanation of the global Schmidt laws observed in galaxies, that the star formation rates correlate with the efficiency of gravitational instability). This paper is organized as follows. In \S~\ref{snmodel}, we describe our models and the relevant parameters. In \S~\ref{snobs}, we describe how synthetic observations are derived. The velocity dispersion dependence on the feedback efficiency, supernova rate and average gas number density is presented and discussed in \S~\ref{snfeed}, \S~\ref{snrate} and \S~\ref{effectn} respectively. Detailed comparison to the observations is performed in \S~\ref{sncomp}. In \S~\ref{snimp} the need for improved numerical models is critically reviewed and in \S~\ref{snconc}, we summarize our results and conclude.

\section{THE MODEL}\label{snmodel}

In order to understand how the SN rate and energy feedback efficiency affect the velocity dispersion of the gas, we resort to a simple numerical model in which the vertical stratification, galactic rotation, magnetic fields and the gas self-gravity are not included. We use the ZEUS-3D code (Stone \& Norman 1992a,b) to solve the equations (mass, momentum and internal energy conservation) of ideal gas dynamics. We simulate a 1 kpc$^{3}$ volume of the ISM with a grid resolution of 128$^{3}$. Periodic boundary conditions are imposed in the three directions. In most simulations and if not specified otherwise, the initial density field is homogeneous with a number density of $\bar{n}=0.5$ cm$^{-3}$. The velocity and temperature are everywhere zero and 10$^{4}$ K, respectively. Radiative cooling of the gas is included by directly interpolating in the the solar metallicity cooling curves of Dalgarno \& McRay (1972) in the temperature range of [100~K,10$^{4}$~K] and of the more recent data of Sutherland \& Dopita (1993) for the temperature range of [10$^{4}$~K-10$^{8.5}$~K]. The gas is not allowed to cool below the minimum temperature $T_{min}=100$ K. The maximum temperatures at the explosion sites reach values of $\simeq 60-70\times 10^{6}$ K which makes the hot gas fall in the hot stable regime. A polytropic equation of state with a specific heat ratio of $5/3$ is used. This proves to be justified because of the lower cutoff temperature of $100$ K that limits over-densities to a few tens of cm$^{-3}$ and therefore the gas remains mostly monoatomic.

 Stellar feedback is modeled as resulting from type II SN explosions only and the energy is injected into the ISM instantaneously. The total energy of each explosion is taken to be $E_{SN}=10^{51}$ erg (Chevalier 1977; Abbott 1982; Woosley \& Weaver 1986 and Heiles 1987). Only a fraction of the total energy is transferred to the ISM in the form of thermal energy. This defines the feedback efficiency parameter $\epsilon$. The energy is injected into $3^{3}$ cells around the central explosion cell following a Gaussian profile. The energy each neighboring cell receives is weighted by the square of its distance to the central cell in order to insure a better isotropy of the explosion. The site where a new SN explosion occurs is chosen randomly under the condition that the local density $n$ is such that $n\ge\bar{n}$. This assumption leads to a more realistic fraction of clustered SN explosions. The time interval between two consecutive SN explosions is given by $\Delta t_{SN}=1/\eta$, where $\eta$ is the SN explosion rate. However, when the time interval between two consecutive SN explosions become shorter than the CFL (Courant-Friedrisch-Levy) time step, more than one SNe are detonated simultaneously at different locations of the grid. The number of SNe detonated in that case is taken to be the closest integer to the ratio $dt/\Delta t_{SN}$, where $dt$ is the CFL time step. We define a Galactic SN rate ($\eta_{G}$) of 2.58 $\times 10^{-4}$ yr$^{-1}$ kpc$^{-3}$ assuming a Galactic radius of 15.5 kpc and a scale height for type II SN of 90 pc (Miller \& Scalo 1979). We use a frequency of 1/57 yr$^{-1}$ (Capellaro, Evans \& Turatto 1999) which is smaller than earlier estimates of 1/50 yr$^{-1}$ by van den Bergh \& McClure (1990) and 1/37 yr$^{-1}$ by Tammann, L\"{o}ffler \& Schr\"{o}der (1994). Note that in our simulations, heating by a UV background radiation is not accounted for. It has not been included because no clear recipe exists that correlates the amount of UV heating that should be added with the different SN rates. One possibility is to distribute a certain fraction of the energy associated with a given SN rate on the total grid. However, at low SN rates, associated with low density environments in the outer parts of galactic disks, the disk might be more easily heated by incident UV photons of extragalactic origin which intensity is not well determined (see discussion in Schaye 2004).

In order to test the effects of numerical resolution, we explode a single supernova in a medium of initial temperature of $10^{4}$ K, initial homogeneous and uniform density of $0.5$ cm$^{-3}$ on a grid representing a physical scale of 1 kpc. The test is similar to the one presented by Mac Low et al. (2005). However the latter authors, aside from having a smaller simulations box (200 pc), assumed the SN remnant to evolve in a medium of negligible pressure (i.e., temperature of 10 K and average density of 0.1 cm$^{-3}$). Fig.~\ref{fig3} displays the time evolution of the shell position R$_{sh}$, measured as being the position of the density peak, in simulations with numerical resolutions of $64^{3}$, the fiducial resolution of $128^{3}$, and $256^{3}$. Clearly, a resolution of $64^{3}$ is not sufficient to resolve properly the dynamics of a single SN remnant. On the other hand, particularly after the first $\sim 0.1$ Myrs in the lifetime of the SN remnant, the discrepancy in the position of the shell at the resolutions of $128^{3}$ and $256^{3}$ is of the order of $10~\%$ and does not exceed the value of $20~\%$. We conclude that our fiducial resolution of $128^{3}$ is good enough for the purpose of studying the global dynamical effects of the energetic input of SN explosions into the ISM.    

\section{ANALYSIS AND DERIVATION OF THE OBSERVABLES}\label{snobs}

The simulations are evolved until kinetic energy reaches a stationary value. Fig.~\ref{fig4} shows the evolution of kinetic energy in a number of simulations. The equilibrium value for the kinetic energy is reached when the dissipation equals the amount of injected kinetic energy. The medium acquires kinetic energy from the SNe explosion-induced thermal pressure gradients, and the thermal pressure gradients associated with TI which occurs in the dense expanding shells. In most simulations, the equilibrium of thermal energy is also reached except for the simulations with a high supernova rate ($\ge 2.5~\eta_{G}$). In the latter simulations, the overlap radii between supernova remnants is very small, and all the gas continues to heat up as more and more energy is injected into the system. Fig.~\ref{fig5} and Fig.~\ref{fig6} show snapshot two-dimensional cuts for models with $(\eta/\eta_{G},\epsilon)=(0.1,0.25)$ and $(1,0.25)$, respectively. In the simulations with the lower rates, larger and denser clouds are able to form under the effect of TI, before being eventually dispersed by a next generation local SN explosion.

 We calculate the velocity dispersion of the gas in two complementary ways. In the first method, we evaluate the characteristic mass weighted velocity $v_{c}$ from the three-dimensional data as 

\begin{equation}
v_{c}= \sqrt{\frac  {\Sigma_{i=1}^{n_{cells}} m_{i} |v_{i}|^{2}}  {\Sigma_{i=1}^{n_{cells}} m_{i}}},  
\label{eq1}
\end{equation}

where the index $i$ runs over the number of cells in the simulation box. An average is made over the last 5 Myrs (5 values) in each simulations in order to smooth for time fluctuations which is particularly useful for the low SN rate simulations. On the other hand, following a more observational approach, the one-dimensional velocity dispersion $\sigma$ is obtained by fitting the mass-weighted line of sight velocity profile. Intensity is assumed to be proportional to the mass along the line of sight. This is particularly true in the case of the \ion{H}{i} line (Rholfs \& Wilson 1996). The velocity profile is then normalized to its maximum value. A second velocity dispersion which we call $\sigma_{\ion{H}{i}}$ is obtained by fitting a velocity profile where only cells which have temperatures $\leq 12000$ K and number densities $n \geq 0.25$ cm$^{-3}$ have been accounted for, thus mimicking the velocity profile of an \ion{H}{i} emission line. We have tested the dependence of the fit parameters on the size of the adopted velocity bin. The relevant parameter (i.e., width of the velocity profile) is practically unchanged as we vary the velocity bin size from 0.1 km s$^{-1}$ up to 2.5 km s$^{-1}$, only the fit-error on the parameters changes, remaining however very close to the value of the spectral resolution. The results we will show correspond to a spectral resolution of 1 km s$^{-1}$. The latter value is characteristic of single-dish radio telescopes (e.g., Effelsberg radio telescope, Green Bank radio telescope) and nears the spectral resolutions obtained with the VLA (Very Large Array) $\simeq 2.5$ km s$^{-1}$ which will be enhanced when the EVLA (Extended Very Large Array) becomes operational. Here also, for each simulation, an averaging over the last 5 Myrs has been performed (5 estimates). Errors on $\sigma$ and $\sigma_{\ion{H}{i}}$ are average values of the individual errors derived from the parameters of the fit functions whereas the error on $v_{c}$ is simply a statistical error over the 5 estimates.

We have attempted to fit the line of sight velocity profiles with several functional distributions, namely, a Gaussian, a Lorentzian, a Moffat profile (modified Gaussian) and a Voigt profile (see Lang 1980 for the mathematical definition of each function). In all cases, the Gaussian, Lorentzian and Moffat fit functions yield the same {\it dispersion} value which has a slight different meaning in each case (see Lang 1980 for details). Fitting with the Voigt profile works only in a limited number of cases, but in those case proves to be a better fit of the profiles. Examples are shown in Figs.~\ref{fig7}-\ref{fig10}. Fig.~\ref{fig7} and Fig.~\ref{fig8} show the 1 km s$^{-1}$ binned total gas and \ion{H}{i} gas line profiles for the model with $(\eta/\eta_{G},\epsilon)=(1,0.25)$, respectively, whereas Fig.~\ref{fig9} and Fig.~\ref{fig10} show the same profiles for the model with $(\eta/\eta_{G},\epsilon)=(0.1,0.25)$. Though Gaussian fitting proves to be quite satisfactory, it is worth mentioning at this stage that our simulated velocity profile have wings that are slightly broader than those associated to Gaussian functions. This non-Gaussianity of the line profiles has been already pointed out by Ricotti \& Ferrara (2002) in their Monte-Carlo models of the ISM dynamics. In a realistic and turbulent ISM, one expects that the turbulent contribution to the line broadening would be described with a Lorentzian and the thermal broadening of the line by a Gaussian. The convolution of the Gaussian and Lorentzian function results in a Voigt profile. However, even when we introduce a thermal broadening component to the line profile (in \S~\ref{sncomp}), the Voigt profile fails to systematically fit the line profiles of the different models.

Since energy injection in our simulations is discrete, a similar value $\eta\times\epsilon$ with different permutations of $\eta$ and $\epsilon$ does not yield similar results because of the differences in the interactions of the expanding shells and non-linear development of TI. Ideally, a full investigation of the two-dimensional parameter space would be necessary, however, this would lead us beyond our current computational capabilities. At this stage, we have taken an intermediate approach and have explored the effects of $\eta$ and $\epsilon$ independently by fixing one parameter and varying the other.

\section{THE EFFECT OF THE FEEDBACK EFFICIENCY}\label{snfeed}

 We performed a first set of simulations for which we varied the supernova feedback efficiency between 0.05 and 1 for a constant value of the supernova rate, $\eta=0.1~\eta_{G}$. Fig.~\ref{fig11} shows the dependence of the characteristic velocity $v_{c}$, line of sight velocity dispersion $\sigma$ and \ion{H}{i} line of sight velocity dispersion $\sigma_{\ion{H}{i}}$ on $\epsilon$. Assuming that all of the three values would be zero if the feedback efficiency is zero, this would mean that $v_{c}$, $\sigma$ and $\sigma_{\ion{H}{i}}$ would rapidly increase with increasing $\epsilon$, followed thereafter by a slower increase at larger values of $\epsilon$. As a velocity has the dimensions of the square root of an energy, we have attempted to fit $v_{c}$, $\sigma$ and $\sigma_{\ion{H}{i}}$ by functions of the form $A \sqrt{\epsilon}$. Fits with these functional forms are over-plotted to the data in Fig.~\ref{fig11}. The values of the fit parameter $A$ are $17.15 \pm 0.59$ km s$^{-1}$, $7.91 \pm 0.36$ km s$^{-1}$ and $6.49 \pm 0.44$ km s$^{-1}$ for $v_{c}$, $\sigma$ and $\sigma_{\ion{H}{i}}$, respectively. In all three curves the simulations are noticeably higher than the estimate of the fit functions for the small values of $\epsilon$. This is plausibly an indication that a second mechanism is responsible for generating kinetic energy in the medium on top of the direct supernova driving in the regime of low energy injection. TI is likely to play an important role in that regime as clouds have time to condense further, thus enhancing the effects of TI, before being destroyed by a local, next generation SN explosion. We will particularly focus on the role of TI in the low energy injection regime in the next section in which we investigate the dependence of the velocity dispersion on the SN rate and for which we have more available simulations.

The total feedback energy (kinetic+thermal) deposited in the ISM by a type II SN exploding in a medium of average density 1 cm$^{-3}$ and solar metallicity was estimated by Thornton et al.(1998) to be $\sim 7 \times 10^{50}$ ergs at the time of the peak luminosity (i.e., $t_{0}$), dropping to $\sim 0.2 \times 10^{50}$ ergs at $t \sim 13 t_{0}$, roughly equally distributed in thermal and kinetic energy. In our models, $t_{0}$ coincides with the time at which the energy is released for each SN explosion and distributed in a volume of radius $\sim 11.7$ pc. However, the estimates of Thornton et al. (1998) should be regarded with some caution essentially because they are based on a one-dimensional model in which the gas has few degrees of freedom. Furthermore, the cooling curve adopted by  Thornton et al. (1998) excludes the radiative cooling from neutral atoms for temperatures lower than 10$^{4}$ K and does not include other potential cooling mechanisms such as neutrino cooling, bremsstrahlung radiation and cooling by thermal conduction. For the investigation of the supernova rate effect on the velocity dispersion, we adopt a conservative value of the feedback efficiency of $\epsilon=0.25$.

\section{THE SUPERNOVA RATE-VELOCITY DISPERSION RELATION}\label{snrate}

The most relevant parameter in our simulations is the supernova rate $\eta$. Fixing $\epsilon$ at a value of 0.25, we performed a set of simulations with different values of $\eta$ ranging from 0.01 to 10 times the Galactic value. The dependence of $v_{c}$ and $\sigma$ on $\eta$ is displayed in Fig.~\ref{fig12} and Fig.~\ref{fig13}, respectively. For $\eta/\eta_{G} \gtrsim 0.5$, Fig.~\ref{fig12} and Fig.~\ref{fig13} show that $v_{c}$ and $\sigma$ increase rapidly with an increasing supernova rate, thus mimicking a starburst regime similar to the one observed in Fig.~\ref{fig2}. A fit to the data for $\eta/\eta_{G}$ between 1 and 10 is $\sigma=0.78(\pm 0.26)~\eta/\eta_{G}+8.30(\pm 1.36)$ km s$^{-1}$. This relation might be useful for semi-analytical modeling of the central regions in galaxies, e.g., the modeling of active galactic nuclei (AGNs), in which hydrostatic equilibrium requires that the effective gravitational potential be balanced by the turbulent pressure of the gas. In current analytical models of AGNs, the intrinsic turbulent velocity of the gas is neglected and gas clouds are assumed to have the same velocity dispersion of the nuclear stellar cluster (e.g., Schartmann et al. 2005). For $\eta/\eta_{G} \lesssim 0.5$, the velocity dispersion shows a slow decrease with a decreasing $\eta$. This transition at $\eta/\eta_{G} \simeq 0.5$ could indicate that the velocity dispersion of the medium in the low rate regime is not fixed by SN driving alone. SNe explosions will cause a certain fraction of the gas to be maintained in the thermally unstable regime when cold gas is restored to the warm phase. Thermal instability occurs in the cooling expanding shells, but also everywhere in the inter-supernova remnants medium where the criterion for thermal instability (see Eq. 9 in V\'{a}zquez-Semadeni, Gazol \& Scalo 2000) is satisfied, thus adding an extra component to the kinetic energy injected into the medium. Since we are using the same cooling curve in the low temperature regime as V\'{a}zquez-Semadeni et al. (2000), the fit used by the latter authors and by Spitzer (1978) for the data of Dalgarno \& McRay (1972) remains valid and the thermally unstable  regime will be confined in the temperature range 398 K $\lesssim$ $T$ $\lesssim$ 10000 K. 

We interpret the flatness of the $\sigma-\eta$ relation for $\eta/\eta_{G} \lesssim 0.5$ as resulting from the interplay between direct supernova driving and thermal instability. Fig.~\ref{fig14} shows, after convergence is reached, the dependence of the volume filling factor of the thermally-unstable gas on the normalized SN rate, whereas Fig.~\ref{fig15} and Fig.~\ref{fig16} display, as examples, the time evolution of the volume filling factor of the unstable (398 K $< T <$ 10000 K), cold ($T < 398$~K) and warm gas ($T > 10000$~K) in the models with ($\eta/\eta_{G},\epsilon$)=($0.01,0.25$) and ($\eta/\eta_{G},\epsilon$)=($0.1,0.25$), respectively. In Fig.~\ref{fig14}, the volume filling factor of the unstable gas which is an indicator of the occurence of TI shows a transition at $\eta/\eta_{G} \simeq 0.5$ which roughly corresponds to the position of the transition observed in the $\eta-v_{c}$ and $\eta-\sigma$ relations, and a non-zero value at smaller values of $\eta$. For the large SN rate values, the gas is predominantly hot, becoming increasingly hotter with time. Therefore the volume filling factor of the unstable gas in that regime is close to zero. In the regime where $\eta/\eta_{G} \lesssim 0.5$, TI is more efficient in converting the gas into the cold phase at the lower end values (i.e., TI is more efficient at $\eta/\eta_{G}=0.01$ than at $\eta/\eta_{G}=0.1$). Fig.~\ref{fig15} and Fig.~\ref{fig16} show that the converged value of the volume filling factor of the cold gas in the simulation with ($\eta/\eta_{G},\epsilon$)=($0.01,0.25$) is $F_{c} \sim 0.8$, whereas this value is $F_{c} \sim 0.5$ for the simulation with ($\eta/\eta_{G},\epsilon$)=($0.1,0.25$). For the same average density in both simulations, SNe exploding in a medium with larger fractions of its volume in the cold phase as in the simulation  with ($\eta/\eta_{G},\epsilon$)=($0.01,0.25$) will evolve in a cooled lower pressure environment than in the simulation with ($\eta/\eta_{G},\epsilon$)=($0.1,0.25$). Numerical simulations of SN driven turbulence by Kim (2004) show that the evolution of SNe remnants in media with lower external pressures leads to higher velocity dispersions of the gas. The existence of a background heating process could modify to some extent the present conclusion. However, the background heating should be strong enough to oppose the dramatic cooling of the medium in the low SN regime and help maintain large fractions of the gas at higher temperatures. However, in the absence of an external heating mechanism to the galaxy, the backgournd heating can only be a small fraction of the energy released by SN explosions and might not play a significant role (see also the discussion in \S~\ref{snimp}). Finally, it is clear that for very small SN rates, SN driving would not be able to sustain any turbulence in the medium. Turbulence will then decay before the next SN explosion occurs.

\section{THE EFFECT OF THE AVERAGE DENSITY}\label{effectn}

As the simulations described in this chapter are not scale free, because of the presence of a realistic cooling function, another relevant point to investigate is the role of the average density. Empirical star formation laws state that the star formation rate decreases with decreasing gas surface density (Schmidt 1959,1963; Kennicutt 1998a,1998b; Dopita \& Ryder 1994; Prantzos \& Silk 1998). In all previous simulations, we have used an average density of 0.5 cm$^{-3}$. It is particularly interesting to test the effect of varying the density for small values of the supernova rate. Over-plotted on Fig.~\ref{fig13} are the results of two simulations with $\eta/\eta_{G}=0.05$ and $\eta/\eta_{G}=0.01$ where the average density have been decreased, by a factor of 5 and 10, respectively. These are shown with the full square and full hexagon, respectively. Dropping the scaling coefficient, if one assumes the supernova rate (i.e., star formation rate)-gas density to follow a Kennicutt type law $\eta/\eta_{G}=\bar{n}^{1.4 \pm 0.15}$ (Kennicutt 1998a,1998b), for an average density value $\bar{n}=0.5$ cm$^{-3}$ corresponds $\eta/\eta_{G}=0.38 \pm _{0.037}^{0.042}$. To values of $\eta/\eta_{G}=0.05$ and 0.01 will correspond, using the same law, average densities of $\bar{n}=0.117 \pm _{0.026}^{0.027}$ cm$^{-3}$ and $\bar{n}=0.037 \pm _{0.0123}^{0.0139}$  cm$^{-3}$, which roughly equal the densities of 0.1 cm$^{-1}$ and 0.05 cm$^{-1}$ we have adopted for those rates. Thus, we can consider the three points ($\eta/\eta_{G}, \bar{n}$[cm$^{-3}]$)=$(0.5,0.5)$, (0.05,0.1) and (0.01,0.05) as being a rough representation of a Kennicutt law in the $\eta-\sigma$ space. Tentatively, the preliminary conclusion we can draw here is that the flatness of the $\sigma-\eta$ relation around $\sim 6$ km s$^{-1}$ can be maintained at low values of the supernova rate if the average density is reduced for lower rates, as predicted by the Kennicutt law. However, more simulations are needed to confirm this result and to probe the results for other star formation laws. Unfortunately, this is beyond the scope of this work, essentially for reasons of CPU time.
                                  
\section{COMPARISON TO THE OBSERVATIONS}\label{sncomp}

In Fig.~\ref{fig17}, the same simulations appearing in Fig.~\ref{fig13} are shown after transforming the SN rate per unit volume into a star formation rate per unit area (units of M$_{\odot}$ yr$^{-1}$ kpc$^{-2}$). We use the transformation of the SN rate into a star formation rate (SFR), $\eta$/SFR=0.0067, derived using the PEGASE stellar population synthesis model (Fioc \& Rocca-Volmerange 1997) and assuming a Salpeter Initial Mass Function (IMF) (Salpeter 1955). We perform a comparison to two galaxies, NGC 628 and NGC 6946 for which the star formation rates have been estimated, at different radii, from H$\alpha$ observations (Martin \& Kennicutt 2001), along with velocity dispersion estimates which are derived from \ion{H}{i} 21 cm line observations (Shostak \& van der Kruit 1984 and Kamphuis \& Sancisi 1993). Fig.~\ref{fig18} shows a comparison of the velocity dispersion measured from the \ion{H}{i} gas (100 K $\lesssim T \lesssim$ 12000 K) velocity profile $\sigma_{\ion{H}{i}}$ to the same observations. A number of remarks can be drawn from the comparisons presented in Fig.~\ref{fig17} and Fig.~\ref{fig18} : (a) The position of the transition to the starburst regime (i.e., location of sharp increase in the velocity dispersion) at around SFR/Area $\simeq 5 \times 10^{-3}-10^{-2}$ M$_{\odot}$ yr$^{-1}$ kpc$^{-2}$ observed in the simulations, is relatively in good agreement with the observations in NGC 628 and NGC 6949. It is also in very good agreement with the transition to the starburst regime observed in Fig.~\ref{fig2}, (b) there is good agreement between our models and the observations at the high SN rate values within $2-3$ km s$^{-1}$. This difference can be easily explained by the effect of beam smearing which tends to increase the observed velocity dispersions, particularly in the inner parts of galaxies  (c) At intermediate and low SN rates, $\sigma$ and $\sigma_{\ion{H}{i}}$ fall below the observed values by a factor of 2-3 even when the density correction related to the Kennicutt law is taken into account in the case of $\sigma_{\ion{H}{i}}$ (open square and star in Fig.~\ref{fig18}). The values of $\sigma$ and $\sigma_{\ion{H}{i}}$ become very similar at the low rates as most of the gas in these simulations has temperatures that are below $12000$ K (see the profiles in Fig.~\ref{fig9} and Fig.~\ref{fig10}). On the other hand no \ion{H}{i} gas is found in the simulations with $\eta/\eta_{G} \geq 1$. It should however be kept in mind that we have adopted a rather conservative value for the supernova feedback efficiency (i.e., $\epsilon=0.25$). Fig.~\ref{fig11} shows that for $\epsilon=0.5$ and 1, for the SN rate value of $\eta/\eta_{G}=0.1$, the velocity dispersions of the \ion{H}{i} gas are of the order $\sim~5$ and $\sim~6$ km s$^{-1}$, respectively, in closer agreement with the observations. 

On the other hand, the velocity dispersions observed in NGC 628 and NGC 6946 (same for the observations presented in Fig.~\ref{fig2}), in addition of being affected to some degree by beam smearing due to the limited resolution of radio telescopes (this is a minor effect at the outer galactic radii), are {\it global} velocity dispersions which do not disentangle the true dynamical value, which is the only one we have in our simulations, from the thermal broadening of the line, simply because the temperature structure of the gas in the observations is not known. Thermal broadening, $v_{T}=(2~k_{b}~T/m)^{1/2}$, is the dispersion of the velocity probability distribution function for an ensemble of particles of a non relativistic and non degenerate gas which is in thermal equilibrium at a kinetic temperature $T$, where $m$ is the mass of the particles and in our case, $m$ is the equal to the proton mass. Seeking a better match between our models and the observations, we have corrected for the effect of thermal broadening in two ways :
   
 (a) {\it {\bf the simplistic method}} : We can subtract, quadratically, from the observational velocity dispersion values, a velocity component associated with thermal broadening at a given equilibrium temperature for the \ion{H}{i} gas, $\sigma_{dyn}=(\sigma_{tot}^{2}-v_{T}^{2})^{1/2}$, where $\sigma_{dyn}$ is the dynamical component that is compared with the dynamical component observed in the simulations and $\sigma_{obs}$ and $v_{T}$ are the observed velocity dispersion and the above mentioned thermal component, respectively. This is obviously a simplification as we assume the \ion{H}{i} gas to have the same temperature at all radii. Fig.~\ref{fig19} shows the corrected velocity dispersions, using this method, for three values of the equilibrium temperature of the \ion{H}{i} gas, namely 100 K, 500 K, and 2000 K. The correction improves the agreement between the observations and the simulations, particularly at the outer galactic radii and if the \ion{H}{i} is assumed to be warmer than 100 K (bottom plots in Fig.~\ref{fig19} where the \ion{H}{i} gas in NGC 628 and NGC 6946 is assumed to be at 2000 K). The fact that the gas in the outer parts could have temperatures higher than 100 K is plausible since the density in the outer parts is low and the gas could be more easily heated by cosmic rays and photoelectric effect. This is unlikely however to be the case for the inner parts of the galaxy where the \ion{H}{i} gas is more likely to have, like in the Milky Way, a non-negligible fraction of the \ion{H}{i} in the cold phase at around 100 K. Nevertheless, recent observations by Heiles (2001) suggest that about half of the mass of the diffuse interstellar gas in the Galaxy may have temperatures which are larger than 100 K (a few hundreds to a few thousands Kelvin). Only an accurate determination of the temperature structure in galaxies such as NGC 628 and NGC 6946 may lay out strong constraints on the contribution of thermal broadening as a function of radius to the total velocity dispersion.

 (b) {\it {\bf the less simplistic method}} : Here, we have assumed that the particles in each cell have a Gaussian velocity profile which is centered around the local dynamical velocity and which have a dispersion in the velocity space equal to $v_{T}$, where $v_{T}$ is the local thermal broadening calculated using the local temperature. The amplitude of the profile is given by the local density. Only cells with $T \leq 12000$ K and $\bar{n} \ge 0.25$ cm$^{-3}$ are taken into account. The individual velocity profiles are summed up in the velocity space and binned with a spectral bin size of 1 km s$^{-1}$. Fig.~\ref{fig20} and Fig.~\ref{fig21} display two mass weighted, thermally broadened velocity profiles corresponding to simulations ($\eta/\eta_{G},\epsilon$)=(1,0.25) and (0.1,0.25), respectively. They can be compared to Fig.~\ref{fig8} and Fig.~\ref{fig10}, respectively, in order to appreciate the effects of thermal broadening on the line profile. Fig.~\ref{fig22} shows a comparison of the \ion{H}{i} velocity dispersion $\sigma_{\ion{H}{i}}$ to the observational data using this more reliable approach for correcting for the effect of thermal broadening. The result is somewhat encouraging. For SFR/Area in the range of $5 \times 10^{-3}-10^{-2}$ M$_{\odot}$ yr$^{-1}$ kpc$^{-2}$ (i.e., $\eta/\eta_{G} \simeq 0.5-1$), the agreement to the observations concerning both NGC 628 and NGC 6946 is quite acceptable. At the lower SN rates values, the agreement is less satisfying even when the Kennicutt rate-adapted average densities are used (open square and star in Fig.~\ref{fig22}). However, a comparison only to the data of NGC 628 and NGC 6946 might be slightly misleading. In the case of galaxies such as NGC 1058 (Dickey, Hanson \& Helou 1990) and NGC 3938 (van der Kruit \& Shostak 1982), for which we unfortunately do not have radially dependent estimates of the star formation rate, the velocity dispersion levels off in the outer radii at a value of the order of 5-6 km s$^{-1}$ which is in better agreement to the values coming out from our simulations.
     
\section{THE NEED FOR IMPROVED MODELS}\label{snimp}

{\it {\bf Effect of the background heating}}~: An important effect, which is not accounted for in our models is a background heating of the gas by the photoelectric effect, cosmic rays and soft X-rays (see also \S 2). These background heating processes might be of little importance in the case of media with SFR/Area $\gtrsim 5\times 10^{-3}$ M$_{\odot}$ yr$^{-1}$ kpc$^{-2}$ (i.e., $\eta/\eta_{G} \gtrsim 0.5$), but might play a significant role in maintaining a warmer phase of the gas at the lower SN rates. Thus, the velocity line profiles might be broader than what we have calculated in the absence of such processes. Kim (2004) showed that the velocity dispersion of the gas in a SN driven medium is reduced if the average pressure of the gas in increased. In the absence of background heating, the gas cools efficiently to the minimum temperature of 100 K in the regions where SN explosions are rare (i.e., this is particularly true for the simulations with the low SN rates). Hence, the gas pressure is reduced and the velocity dispersion enhanced. On the other hand, the shock produced by a SN explosion expanding in a cold medium would lead to higher compressions than a shock propagating in a warmer medium. The stronger compressions would lead to an enhanced cooling in the compressed SN shells which causes the energetic content of the explosion to be depleted faster. The intensity of the background heating is difficult to estimate for systems with different SN rates. As no established formulations of this problem exist in the literature, we intend, in future work, to model the background heating as being a fraction of the total SN heating and quantify its effects on the resulting velocity dispersion of the gas. 
   
{\it {\bf Effect of the vertical structure}}~: The vertical stratification, which we have neglected in this work, might be one of the physical effects that we need to include first in subsequent models. We intend to perform models with a much larger length of the box in the vertical direction, using outflow boundary conditions to allow the gas to escape from the upper and lower boundaries of the simulation box. The escaping hot gas would not affect the \ion{H}{i} 21 cm line profile. However, denser blobs of gas which are expelled at higher latitudes by the SN explosions will cool and fall back into the disk in the form of high velocity clouds. In principle, these high velocity clouds should be observed as broad wings in the velocity profile and are usually fitted with a second, broader Gaussian function. However, in-falling decelerated gas, close to the galactic disk will be most probably mixed with the local gas in the disk, thereby contributing, eventually, to the broadening of the velocity profile. The impact of HVCs on the galactic can also substantially enhance the local level of turbulence particularly for massive HVCs. 

{\it {\bf Effect of the chemistry}}~: In the simulations presented in this work, the cooling curve we have used is a solar metallicity curve that assumes chemical equilibrium. This cooling curve describes principally the radiative cooling by neutral atoms whereas molecular cooling by molecules such as the H$_{2}$ molecule is neglected. The additional cooling at lower temperatures might enhance the local effects of TI and increase the value of the velocity dispersion. At the high and intermediate SN rates, this might not be of much relevance as the over-densities that are produced are a factor 3-5 the average density (a few cm$^{-3}$) (see Fig.~\ref{fig5}) and no dense molecular material is expected to form. At low rates, clouds have time to form and condense further before being destroyed by the next generation of SNe (see Fig.~\ref{fig6}). Molecular hydrogen, starts to form when densities of the order of $10^{3}$ cm$^{-3}$ are reached and which become shielded against UV radiation (Bergin et al. 2004). Such densities are not reached in our simulations, essentially because of the limitations in the numerical resolution. The introduction of a simple chemical network to follow locally the fraction of molecular hydrogen in a hydrodynamical adaptative-mesh-refinement (AMR) code would help tackle the problem of molecules formation in the expanding shells more accurately, and help better account for the additional molecular cooling and its effects on the velocity field.

{\it {\bf Effect of the metallicity}}~: The occurrence and efficiency of the thermal instability is intimately related to the shape of the cooling curve which characterizes the medium. Cooling curves are a reflection of the strength of the emission lines of atoms present in the medium. At lower metallicity, the emission lines from metals are weaker and the cooling less efficient. The dependence of the cooling rate on the metallicity has been calculated by Boehringer \& Hensler (1989), unfortunately only in the temperature range $10^{4}-10^{8}$ K. Their results show that the cooling rates may differ, in some temperature ranges, by several orders of magnitude for metallicities between $10^{-2} \leq Z/Z_{\odot} \leq 2 $ ($Z/Z_{\odot} =2$ is the upper metallicity limit in their calculations) and becomes independent of the metallicity for $Z/Z_{\odot} < 10^{-2}$. In future work, we plan to investigate the role of metallicity on the dynamics of the ISM by including it as a parameter in our simulations.

{\it {\bf Effect of self-gravity}}~: Kim et al. (2003) in some of their MRI simulations which include gravity show that for a single-phase medium and for a Toomre parameter $Q_{th} \ge 1.7$ which is more appropriate for the external regions of galactic disks where the gas surface density is low, self-gravity can be responsible for only 20 $\%$ of the level of $1.6-3.2$ km s$^{-1}$ turbulence generated by the MRI. However, in our models with SN explosions, the presence of self-gravity, provided enough numerical resolution is affordable to resolve the expanding shells, might lead to the development of Rayleigh-Taylor and Kelvin-Helmotz instabilities (Shu 1992), particularly in the case of SN remnants evolving in non-spherical environments (see e.g., Gazol-Pati\~{n}o \& Passot 1999).

{\it {\bf Effect of the magnetic field}}~: In the present work, magnetic fields have been neglected. Several authors have performed simulations of a supernova explosions in a magnetized medium (e.g., Ferri\`{e}rre 1991; Gazol-Pati\~{n}o \& Passot 1999; Korpi et al. 1999a,b; Kim 2004; de Avillez \& Breitschwerdt 2005; Mac Low et al. 2005). The latter simulations agree that the effect of magnetic fields in essentially to oppose the radial expansion of a supernova remnant, thus reducing the energy transmitted to the ISM and reducing the velocity dispersion. In particular,  the results of Kim (2004) show that the velocity dispersion in a SN driven medium is related to the total (thermal + magnetic) pressure. Dispersions are smaller in environments with higher total pressures (i.e., higher magnetic field values for a given thermal pressure). For a comparison, at the Galactic SN rate we find that the one-dimensional total gas velocity dispersion is $\sim 8.5$ km s$^{-1}$, whereas in the magnetized case Kim (2004) finds $\sim 9$ and $\sim 10$ km s$^{-1}$ for a weak magnetic field value of 2 $\mu$G associated to an average density of 0.2 cm$^{-3}$ in the directions parallel to the mean field and perpendicular to it, respectively. For a stronger field value of 8 $\mu$G associated to an average density of $0.8$ cm$^{-3}$, Kim (2004) finds velocity dispersions of $\sim 5$ and $\sim 7$ km s$^{-1}$ in the directions parallel and perpendicular to the field, respectively. In view of Kim's (2004) results, we speculate that in our simulations, as SNe occur in region of higher density, the existence of a magnetic field which would be compressed in those regions, would have the effect of lowering our measured values of the total gas velocity dispersion by a factor of a few tens of percent (typically 10-20 $\%$).  

\section{SUMMARY AND DISCUSSION}\label{snconc}

In this paper, we investigated the dependence of the velocity dispersion in the interstellar medium (ISM) on the supernova (SN) rate $\eta$, the SN feedback efficiency $\epsilon$, and in some cases on the ISM average number density $\bar{n}$. We use local, three-dimensional numerical simulations in which SN type II explosions are detonated in random positions of the grid, separated by time intervals which are inversely proportional to the SN rate. Radiative cooling of the gas is also taken into account with a minimum cutoff temperature of 100 K. For the purpose of simplifying the problem, other physical processes and characteristics of galactic disks such as the vertical stratification, magnetic fields and gravity are neglected. For each simulation, we calculate the three-dimensional characteristic velocity dispersion $v_{c}$ (Eq.~\ref{eq1}) and the one-dimensional velocity dispersion, $\sigma$, obtained by fitting the line of sight velocity profile with a Gaussian function. We also calculate $\sigma_{\ion{H}{i}}$, which is the one-dimensional velocity dispersion obtained from a Gaussian fit of the line of sight velocity profile of the gas with a temperature $T \leq 12000$ K and a number density $n \geq 0.25$ cm$^{-3}$ (i.e., the \ion{H}{i} gas). Ideally, a full investigation of the two-dimensional parameter space ($\eta,\epsilon$) would be necessary, however, this would lead us beyond our current computational capabilities. At this stage, we have taken an intermediate approach and explored the effects of $\eta$ and $\epsilon$ independently by fixing one parameter and varying the other.

Our results show that $v_{c}$, $\sigma$ and $\sigma_{\ion{H}{i}}$ depend on the SN feedback efficiency $\epsilon$ as $A \sqrt{\epsilon}$, where $A$ is a scaling coefficient, different for each quantity. This is expected as the velocity has the dimensions of the square root of an energy. In a second set of simulations, we fixed $\epsilon$ to a value of 0.25 and varied the SN rate in the range [0.01,10] $\eta_{G}$, where $\eta_{G}$ is the Galactic SN rate. We compare the velocity dispersion for the simulations with the different SN rates to NGC 628 and NGC 6964 for which both the velocity dispersion and the the star formation rates are radially resolved. The dependence of the velocity dispersion on the SN rate is complex. For values of $\eta/\eta_{G} \gtrsim 0.5$ (i.e., SFR/Area $\gtrsim 10^{-2}$ M$_{\odot}$ yr$^{-1}$ kpc$^{-2}$), $v_{c}$ and $\sigma$ increase sharply with increasing values of $\eta$. For $\eta/\eta_{G} \lesssim 0.5$, $v_{c}$ and $\sigma$ show a slower decrease with a decreasing $\eta$. This transition at $\eta/\eta_{G} \simeq 0.5$ is probably an indication that the velocity dispersion of the medium in the low rate regime is not fixed by SN driving alone. We interpret the flatness of the $\sigma-\eta$ relation as resulting from the efficient development of thermal instability (TI) in the medium with the low SN rates. We quantify the efficiency of TI by evaluating the volume filling factor of the unstable gas $F_{u}$ for the different SN rates. $F_{u}$ appears to be correlated with the velocity dispersion. For large values of the SN rate $\eta/\eta_{G} \gtrsim 0.5-1$, $F_{u}$ is close to zero, with a transition to non zero values at $\eta/\eta_{G} \sim 0.5$, peaks at $\eta/\eta_{G} \sim 0.1$, before decreasing slowly at smaller SN rates. Interestingly, the position of the transition to the starburst regime (high SFR rates associated with high velocity dispersions) is in relatively good agreement with the one seen in the observations (see Fig.~\ref{fig2}, Fig.~\ref{fig18} and Fig.~\ref{fig19}).

\begin{figure}
\plotone{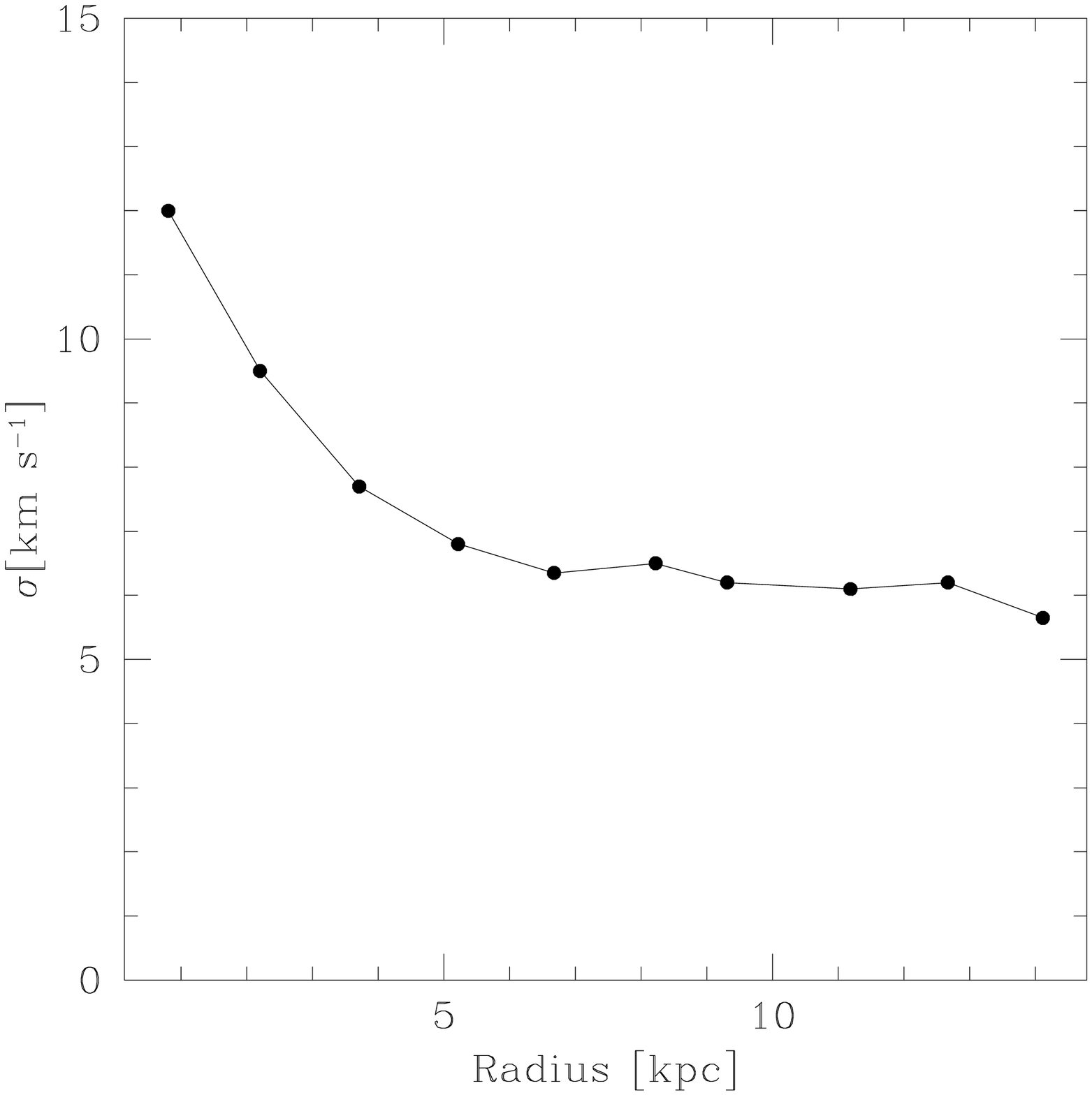} 
\caption{Radial dependence of the velocity dispersion in NGC 1058. Adapted from Dickey, Hanson \& Helou (1990) (Fig. 3 in their paper), assuming a distance of 10.2 Mpc (Boroson 1981).}
\label{fig1}  
\end{figure}

\begin{figure}
\plotone{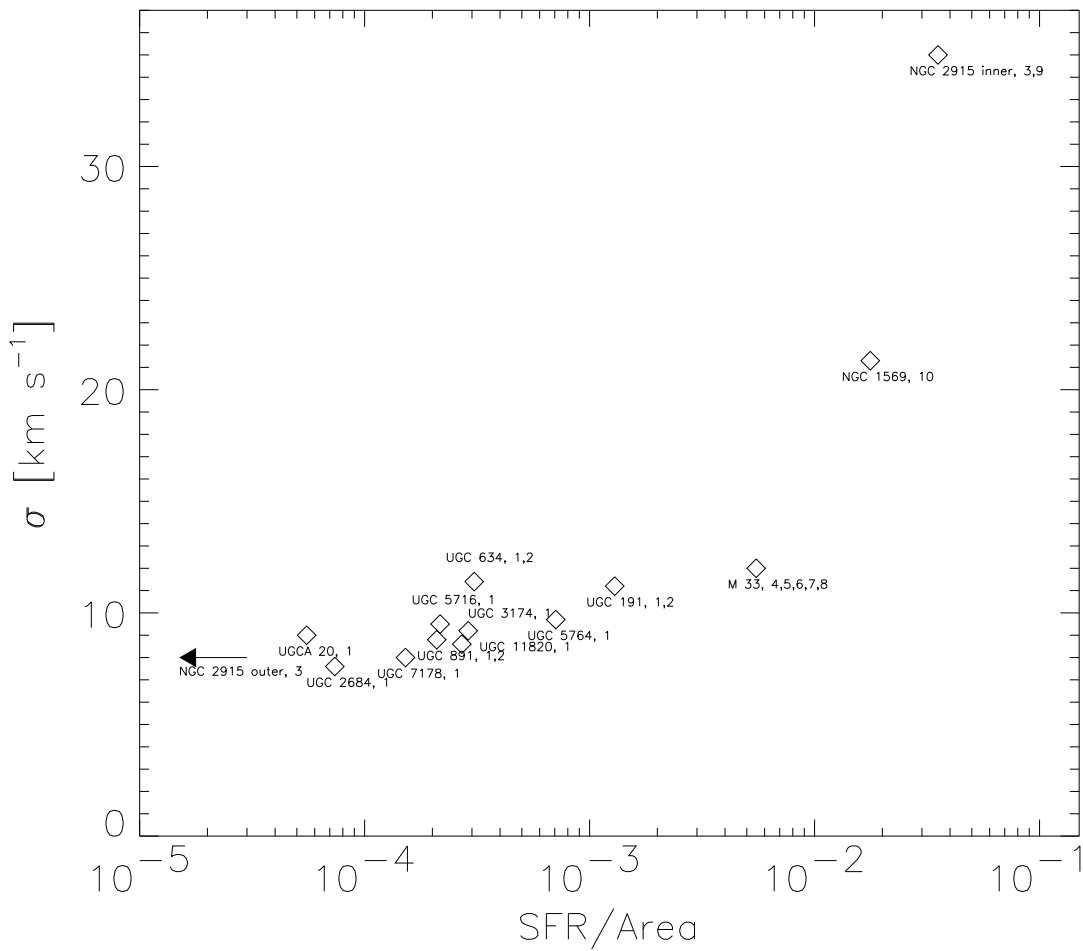} 
\caption{Characteristic velocity dispersion $\sigma$ of a sample of galaxies as a function of the surface averaged star formation rate (units of M$_{\odot}$ yr$^{-1}$ kpc$^{-2}$). The latter value is calculated from H$_{\alpha}$ observations. The galactic surface is calculated as being $\pi~ (3~r_{d})^{2}$, where $r_{d}$ is the radial length scale of each galaxy. References are (1) van Zee et al. (1997) (2) van Zee (2001) (3) Meurer et al. (1996) (4) Hutchmeier (1973) (5) Gordon (1971) (6) de Jager \& Davies (1971) (7) Hippelein et al. (2003) (8) Elmegreen \& Elmegreen (1984) (9) Meurer, Mackie \& Carignan (1994) (10) Stil \& Israel (2002).}
\label{fig2}  
\end{figure}

\begin{figure}
\plotone{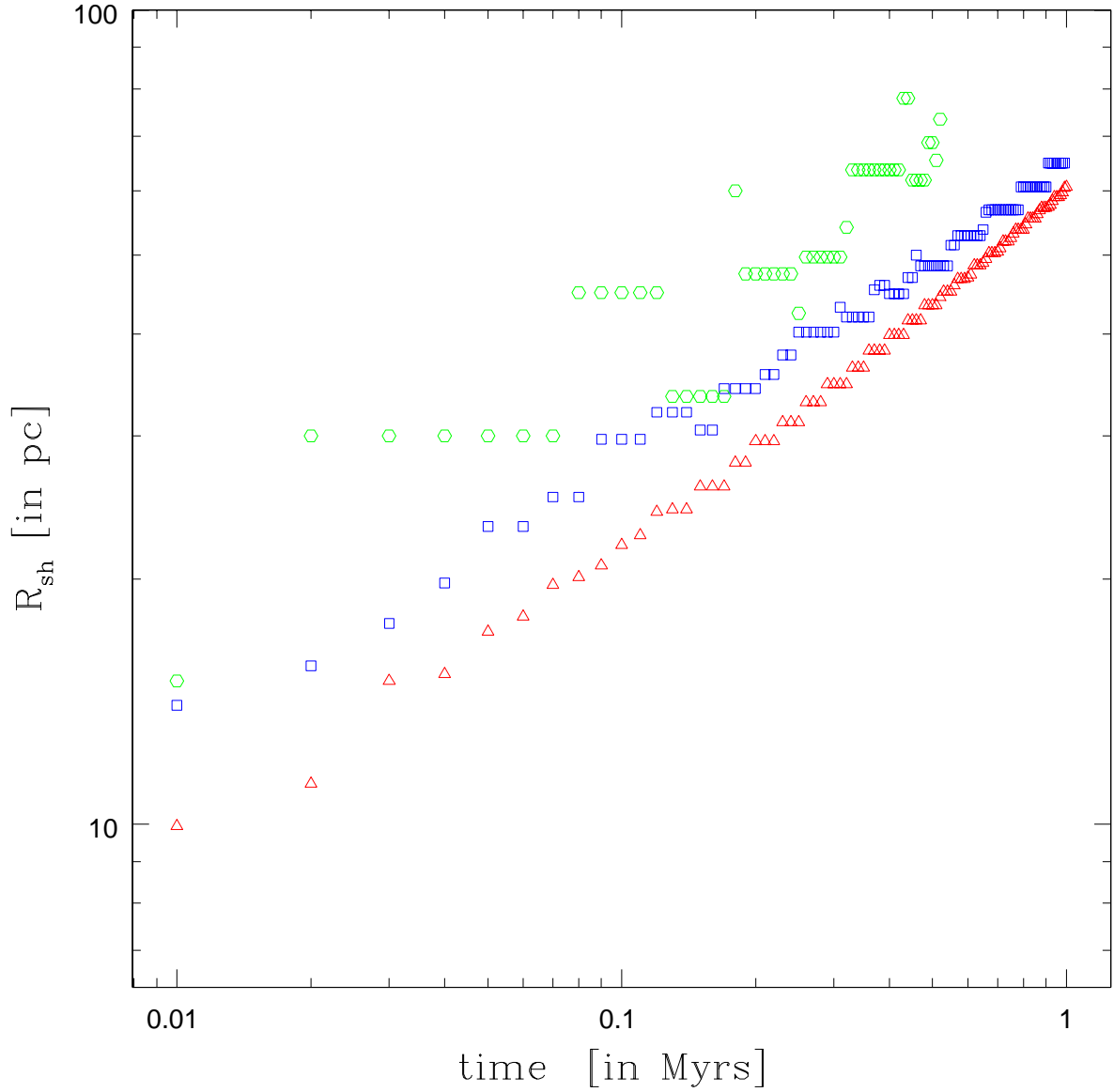} 
\caption{Convergence study of the shell radius for a supernova explosion occurring in medium with initial temperature of $10^{4}$ K and uniform and homogeneous density of 0.5 cm$^{-3}$. The size of the box is 1 kpc. The three resolutions correspond to 64$^{3}$ (green, hexagons), 128$^{3}$ (blue, squares), and $256^{3}$ (red, triangles).}
\label{fig3}  
\end{figure}

\begin{figure}
\plotone{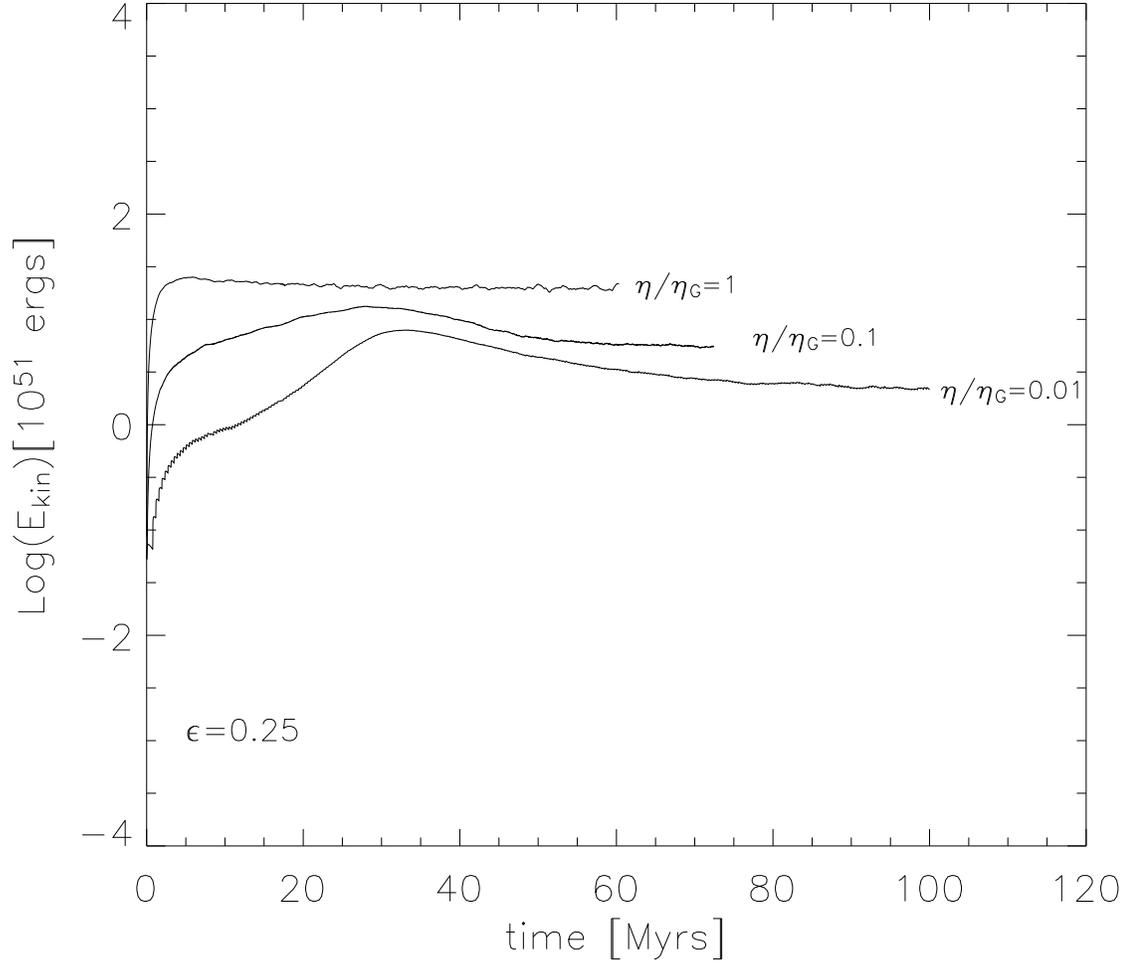} 
\caption{Time evolution of the kinetic energy in a number of selected simulations. The supernova rate, normalized to the Galactic value is shown for each curve. The feedback efficiency for all three models is $\epsilon=0.25$.}
\label{fig4}  
\end{figure}

\begin{figure}
\plotone{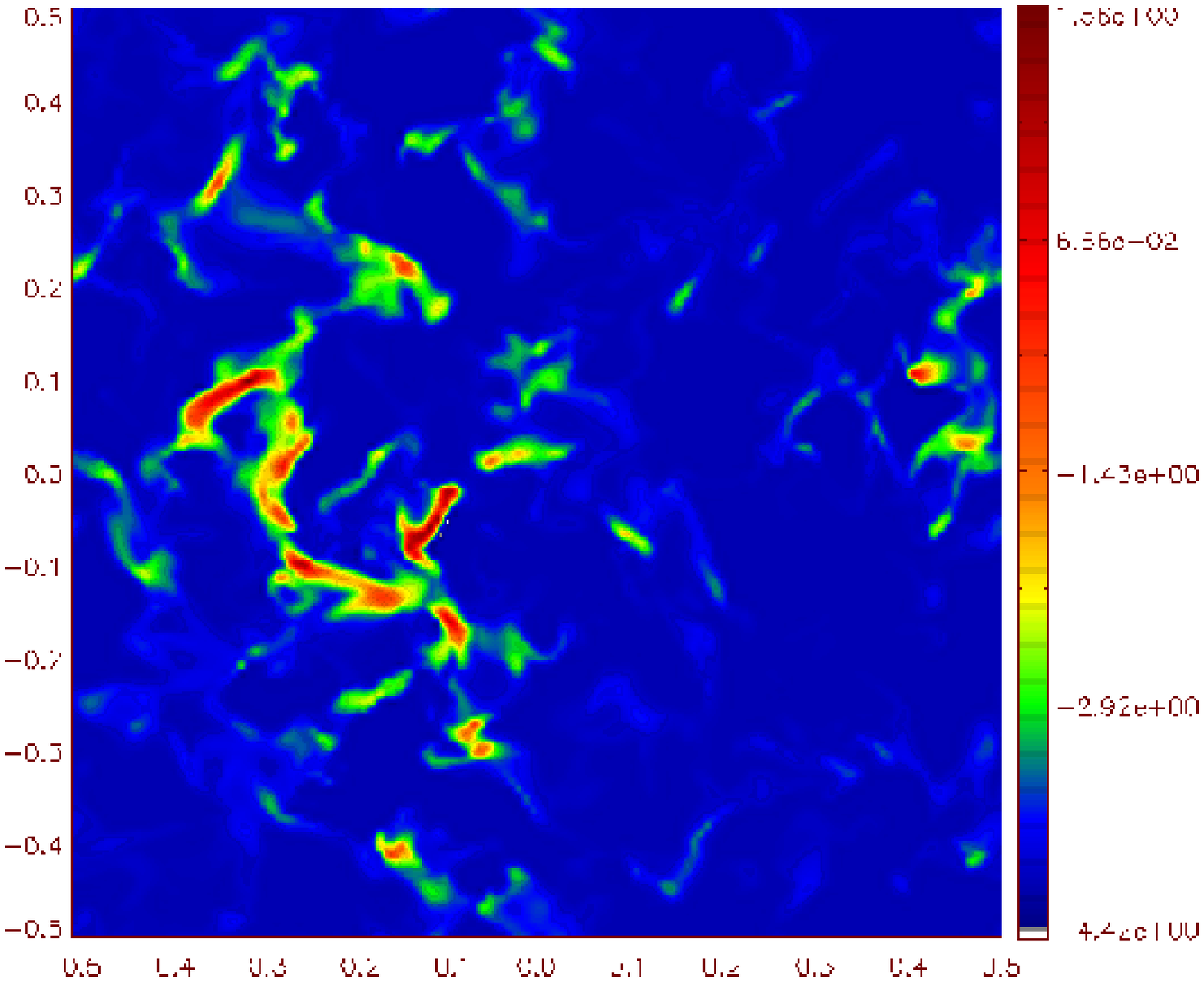} 
\caption{Density cut in the mid-plane of the data cube at time=72 Myrs for the model with $(\eta/\eta_{G},\epsilon)=(0.1,0.25)$. The color bar data scale is logarithmic.}
\label{fig5}  
\end{figure}

\begin{figure}
\plotone{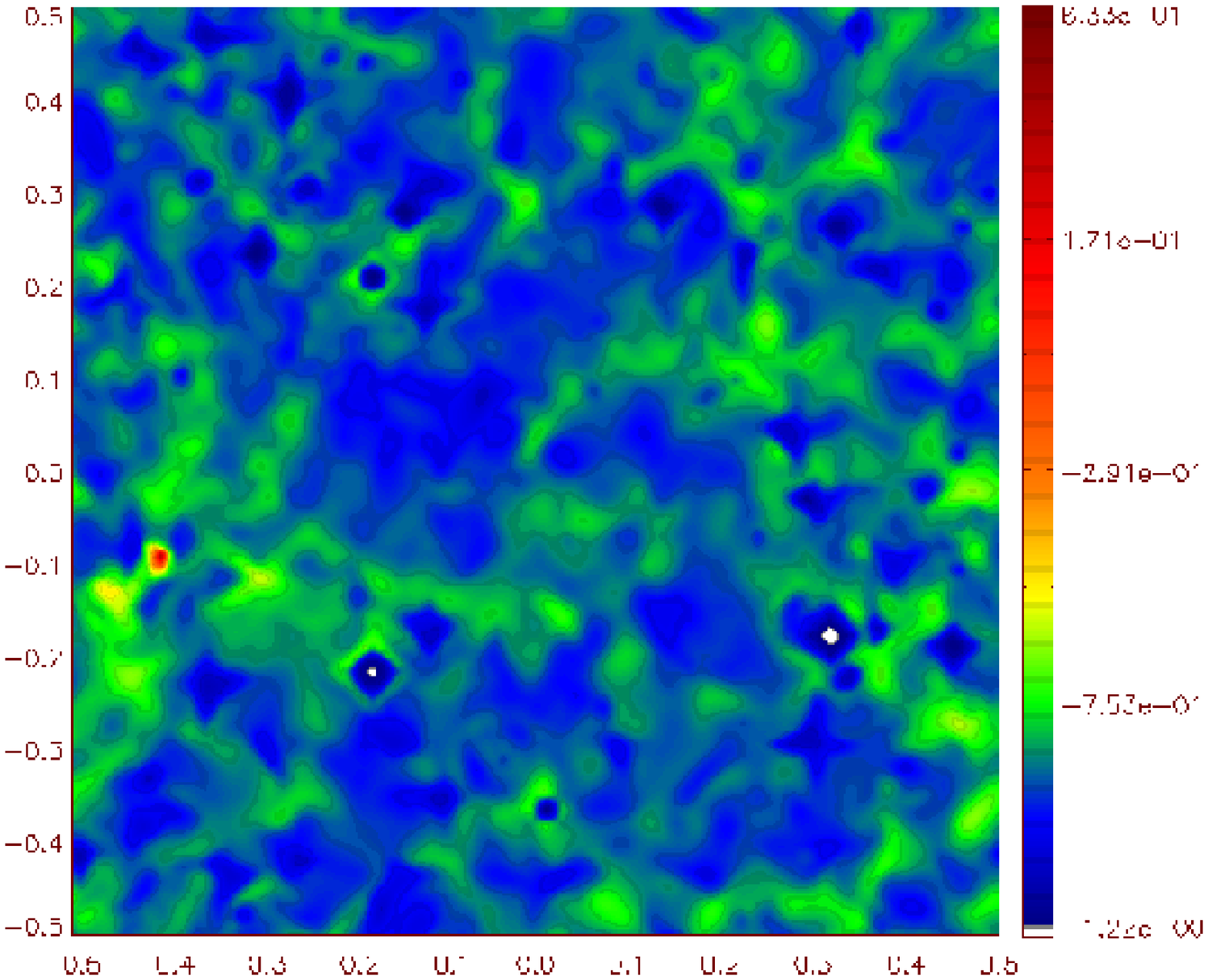} 
\caption{Density cut in the mid-plane of the data cube at time=60 Myrs for the model with $(\eta/\eta_{G},\epsilon)=(1,0.25)$. The color bar data scale is logarithmic.}
\label{fig6}  
\end{figure}

\begin{figure}
\plotone{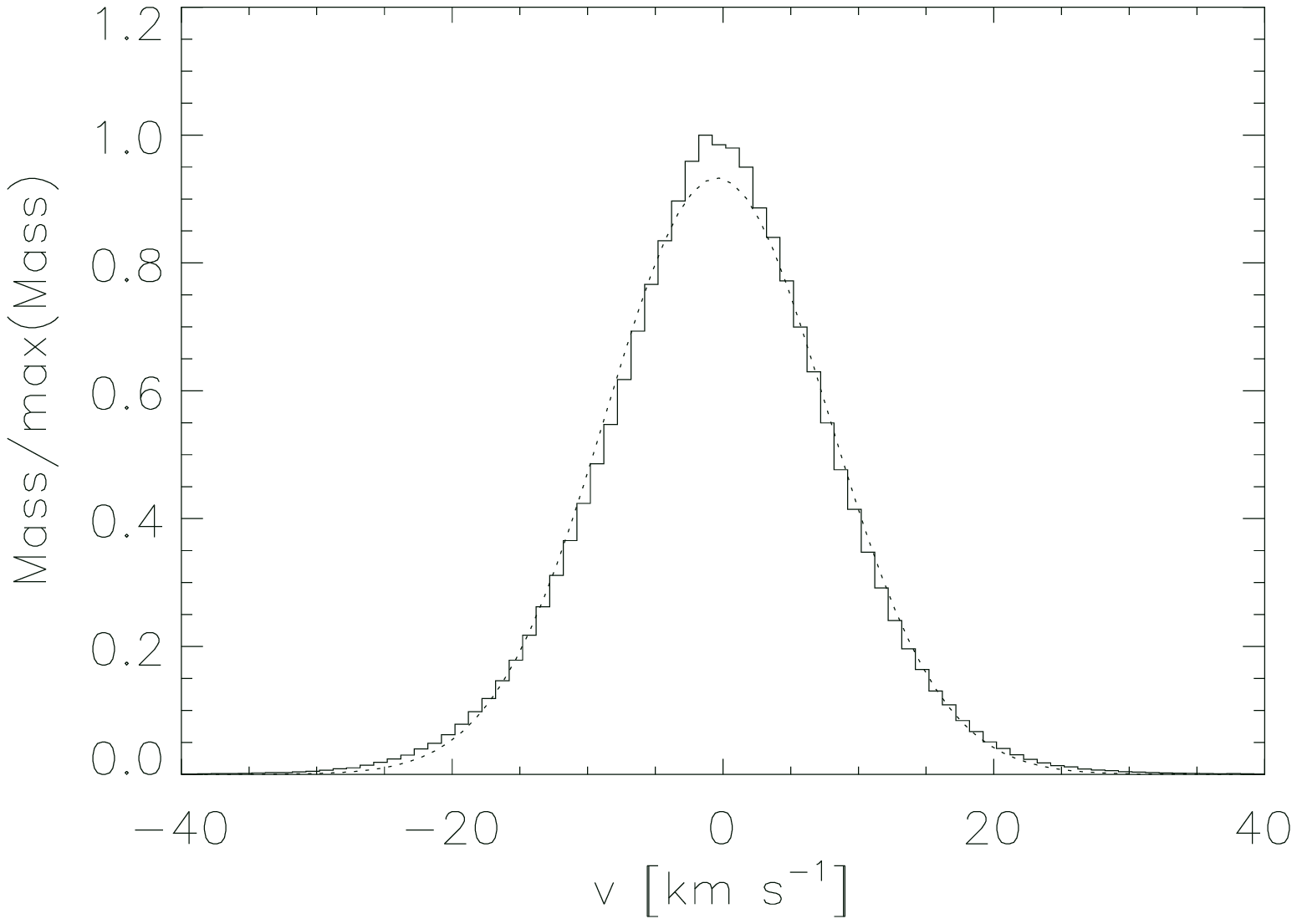} 
\caption{Mass weighted velocity profile for the model with ($\eta/\eta_{G},\epsilon$)=(1,0.25) and a spectral bin size of 1 km s$^{-1}$. The profile samples all gas elements in the simulation box. Over-plotted is the Gaussian fit (dotted line).}
\label{fig7}  
\end{figure}

\begin{figure}
\plotone{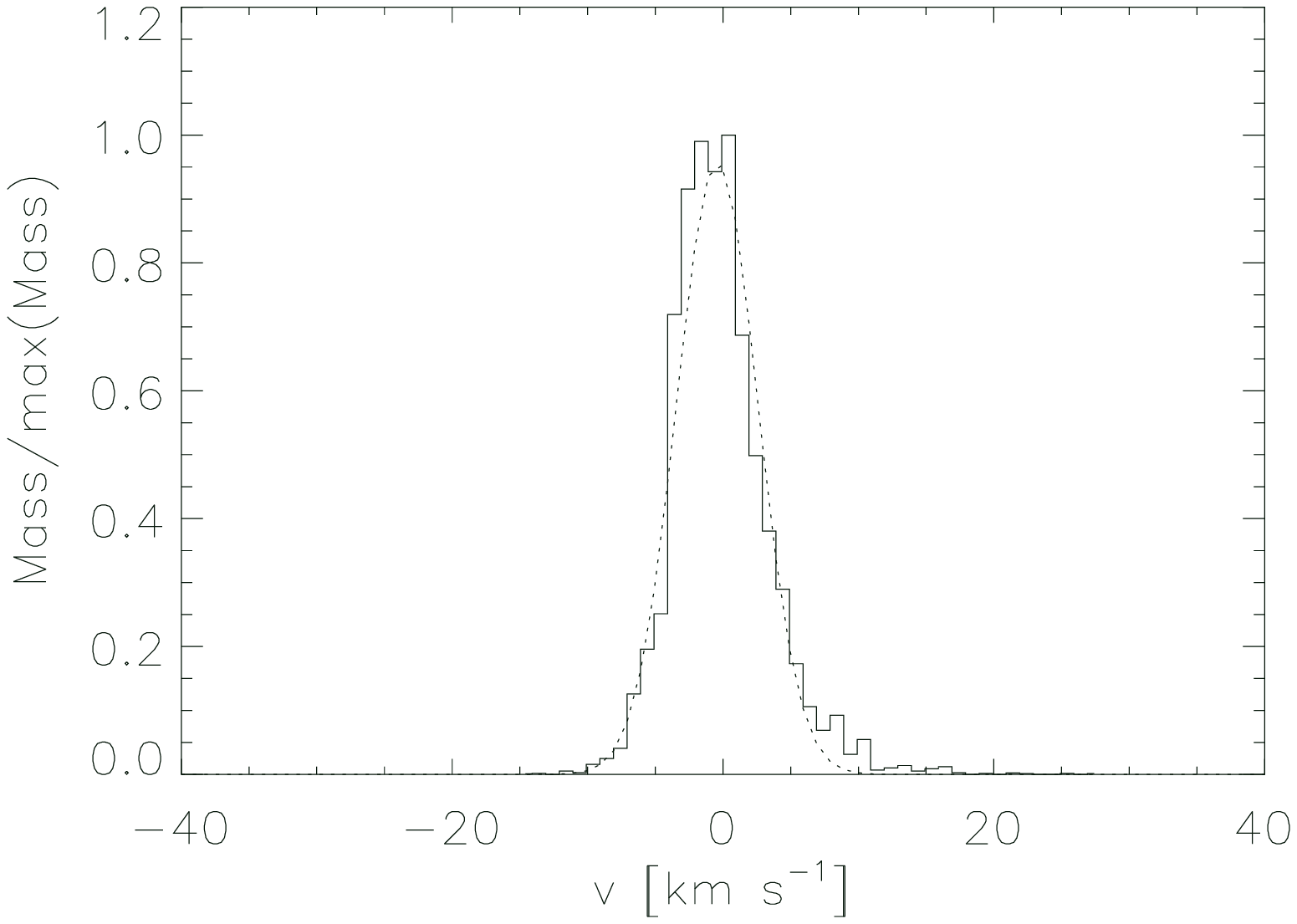} 
\caption{Mass weighted velocity profile for the model with ($\eta/\eta_{G},\epsilon$)=(1,0.25) and a spectral bin size of 1 km s$^{-1}$. The profile samples only gas elements with $T \leq 12000$ K and $n \geq 0.25$ cm$^{-3}$. Over-plotted is the Gaussian fit (dotted line).}
\label{fig8}  
\end{figure}

\begin{figure}
\plotone{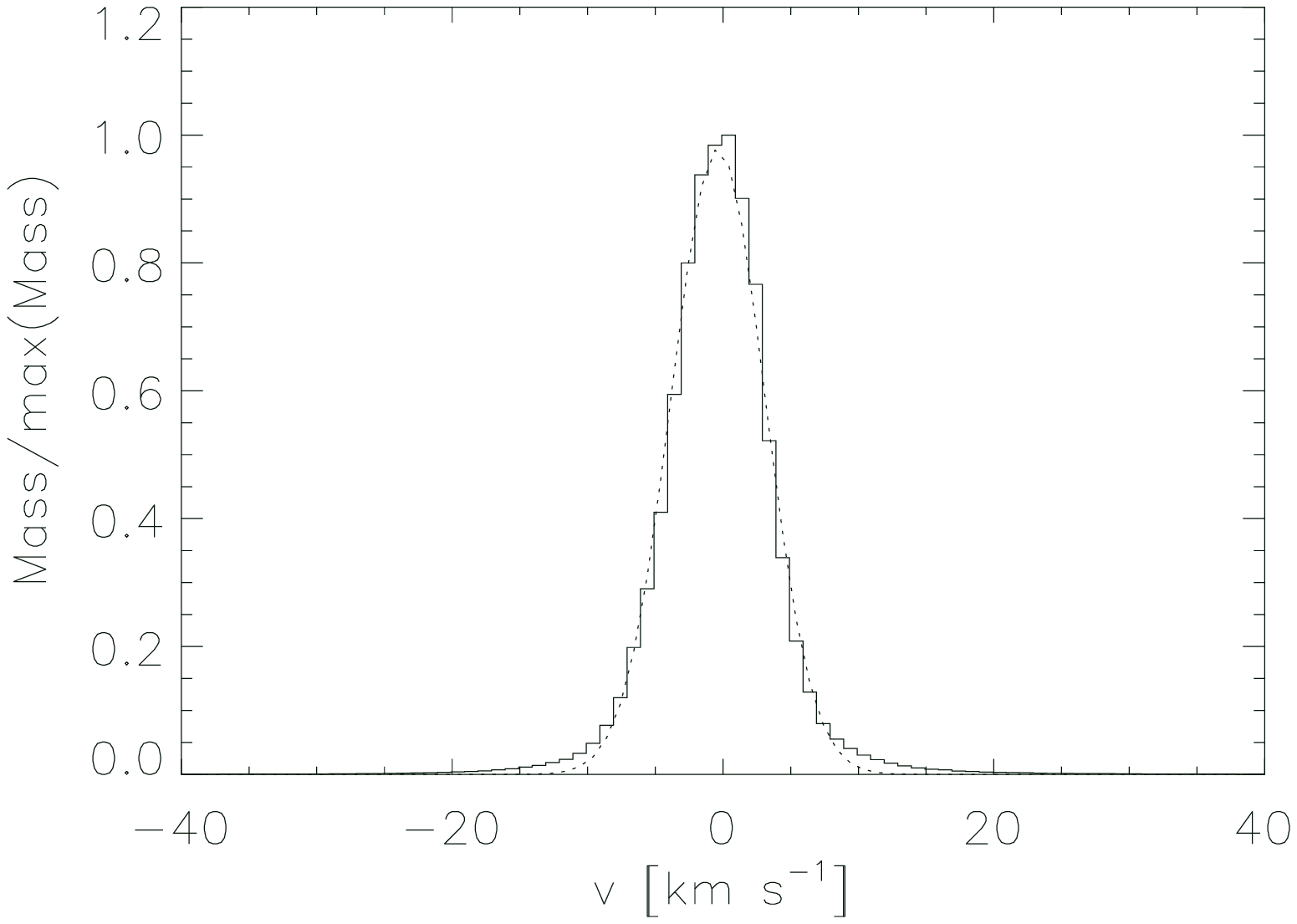} 
\caption{Mass weighted velocity profile for the model with ($\eta/\eta_{G},\epsilon$)=(0.1,0.25) and a spectral bin size of 1 km s$^{-1}$. The profile samples all gas elements in the simulation box. Over-plotted is the Gaussian fit (dotted line).}
\label{fig9} 
\end{figure}

\begin{figure}
\plotone{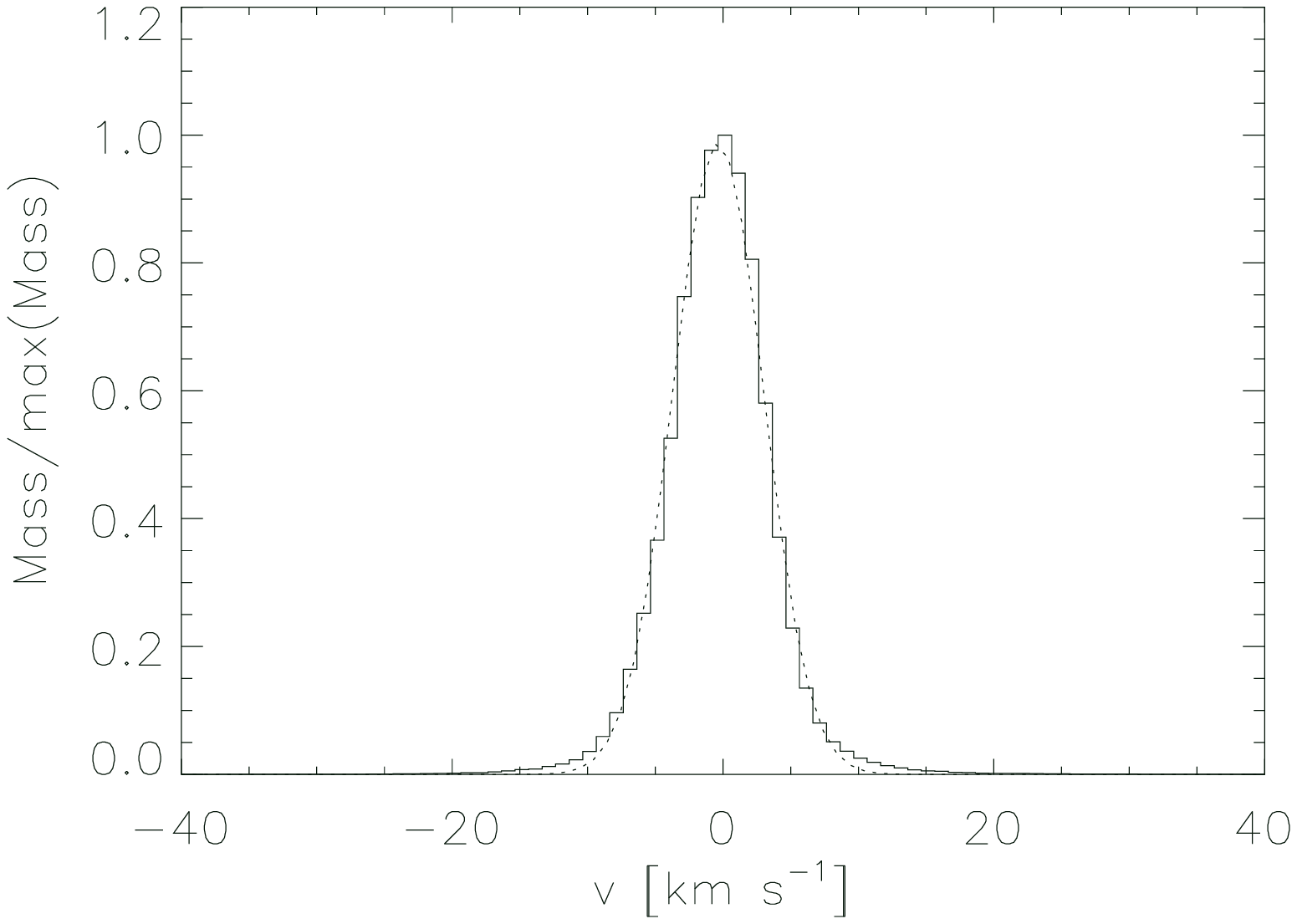} 
\caption{Mass weighted velocity profile for the model with ($\eta/\eta_{G},\epsilon$)=(0.1,0.25) and a spectral bin size of 1 km s$^{-3}$. The profile samples only gas elements with $T \leq 12000$ K and $n \geq 0.25$ cm$^{-3}$. Over-plotted is the Gaussian fit (dotted line).}
\label{fig10}  
\end{figure}

\begin{figure}
\plotone{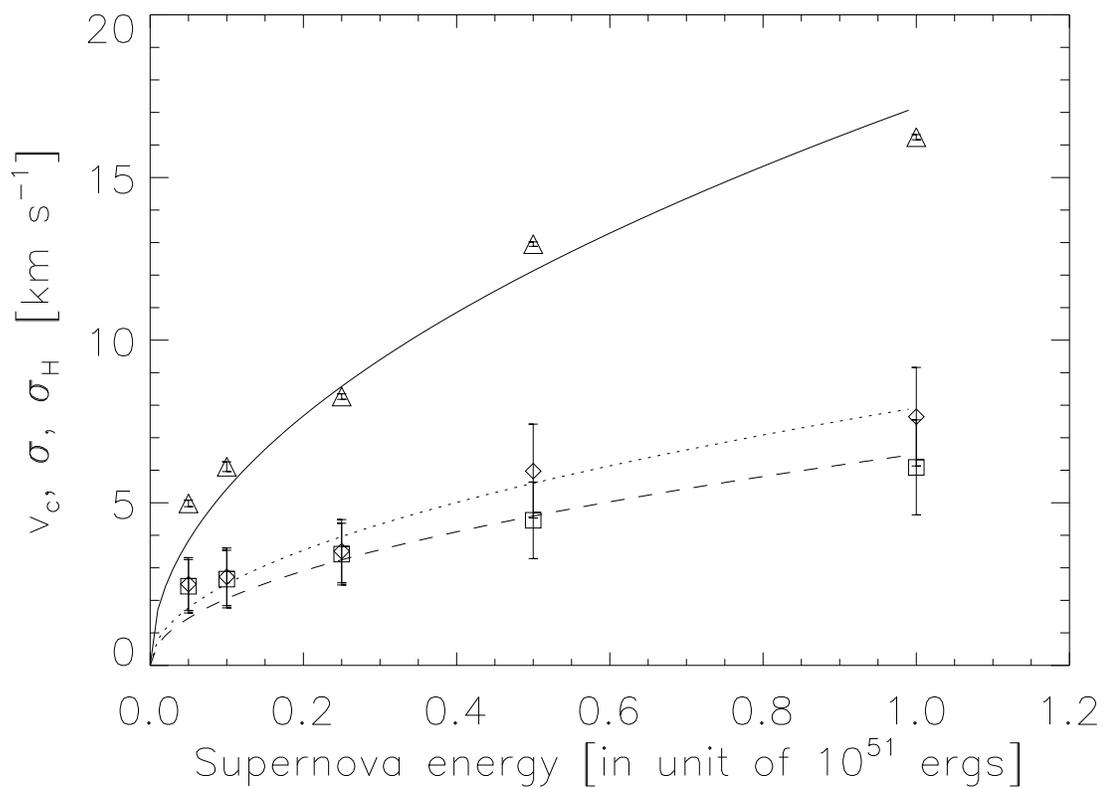} 
\caption{Characteristic velocity $v_{c}$ (Eq.~\ref{eq1}) (triangles), mass-weighted line of sight velocity $\sigma$ (diamonds) and \ion{H}{i} gas mass-weighted line of sight velocity $\sigma_{\ion{H}{i}}$ (squares) (both have spectral bin size of 0.1 km s$^{-1}$) as a function of the supernova feedback energy equal to $\epsilon \times 10^{51}$ ergs.}
\label{fig11} 
\end{figure}

\begin{figure}
\plotone{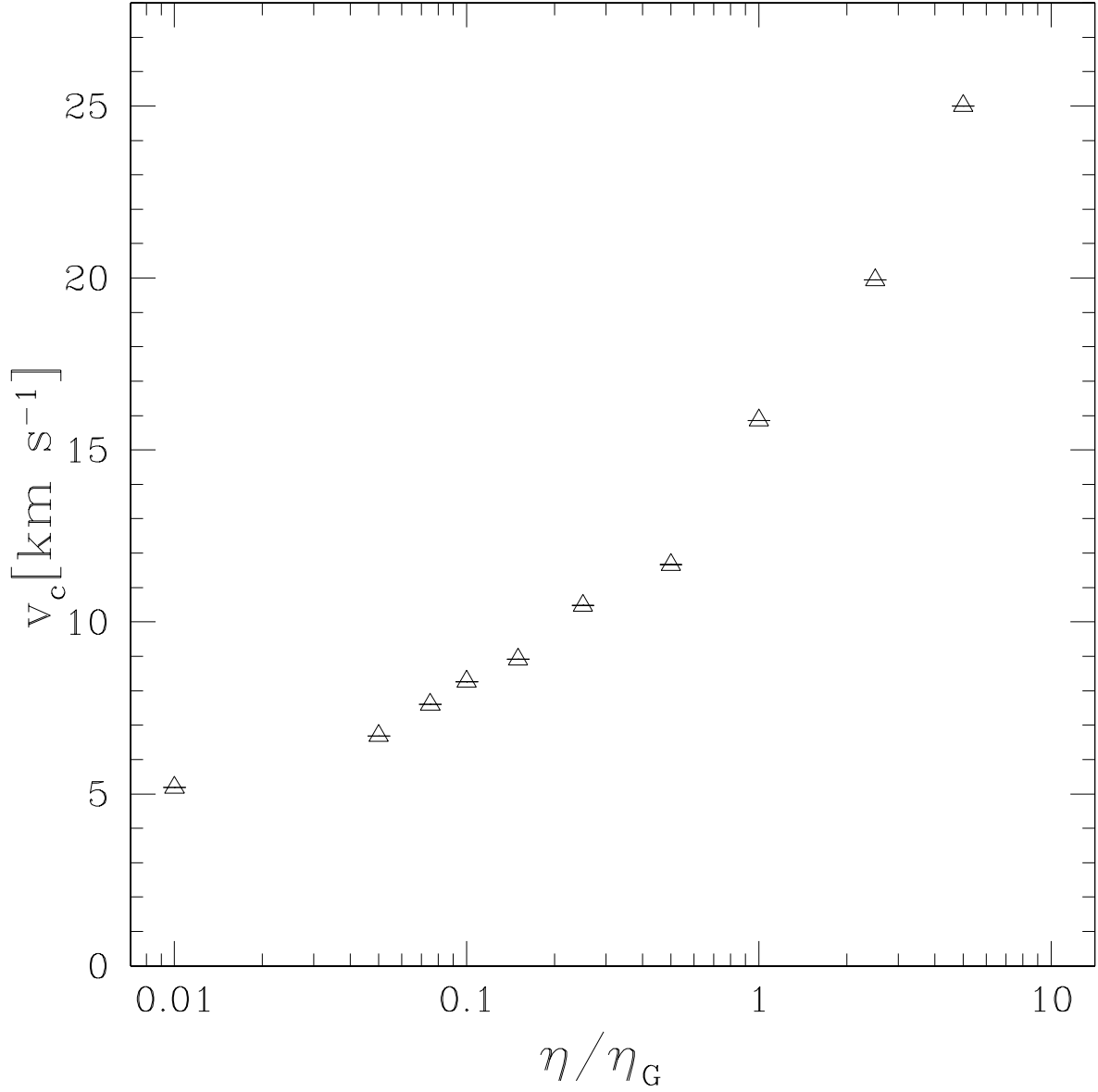} 
\caption{Characteristic velocity $v_{c}$ as a function of supernova rate $\eta$ (normalized to the Galactic value $\eta_{G}$).}
\label{fig12}  
\end{figure}

\begin{figure}
\plotone{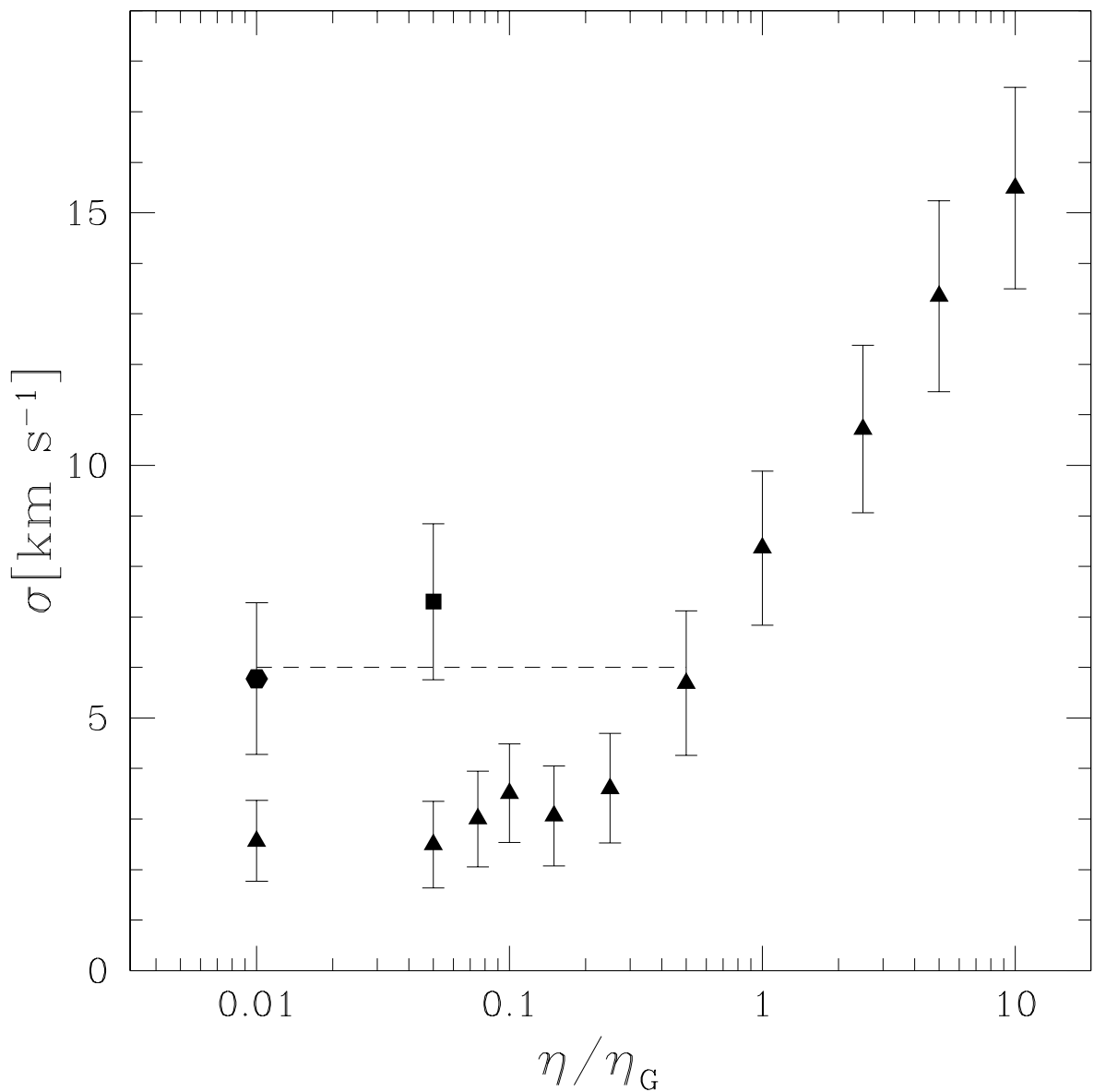} 
\caption{Velocity dispersion of the gas $\sigma$ as a function of the supernova rate $\eta$ (normalized to the Galactic value $\eta_{G}$, filled triangles). Filled square and filled hexagon correspond to simulations where the average density has been reduced by factors of 5 and 10, respectively. The displayed data has been obtained using a spectral bin size of 1 km s$^{-1}$. The constant line at 6 km s$^{-1}$ is drawn to guide the eye.}
\label{fig13}  
\end{figure}

\begin{figure}
\plotone{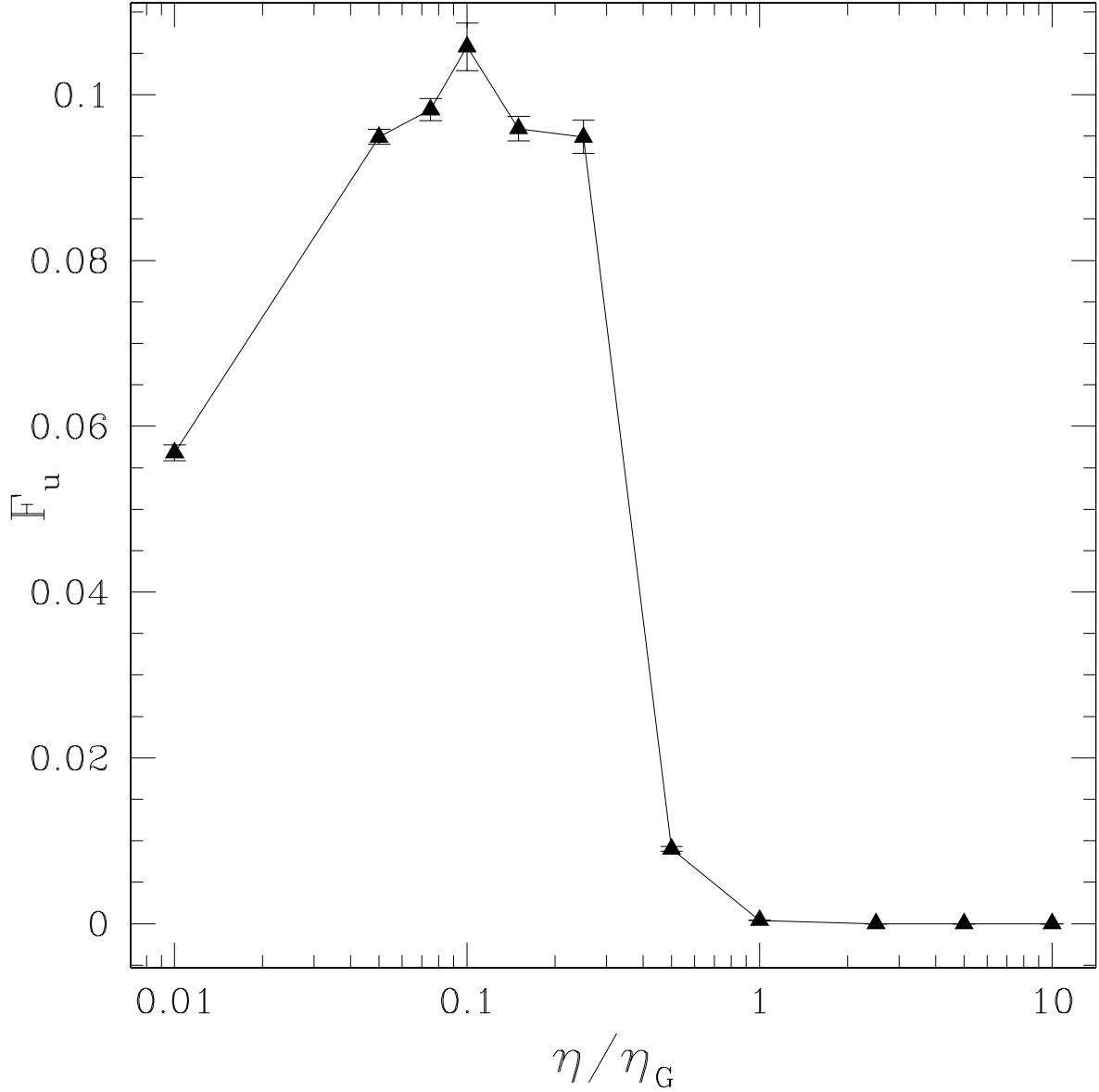} 
\caption{Volume filling factor of the unstable gas (398 K $\lesssim$ T $\lesssim$ 10000 K) as a function of the supernova rate $\eta$ (normalized to the Galactic value $\eta_{G}$). The error bars represent the statistical errors for 5 estimates of each value (last 5 Myrs in each simulation).}
\label{fig14}  
\end{figure}

\begin{figure}
\plotone{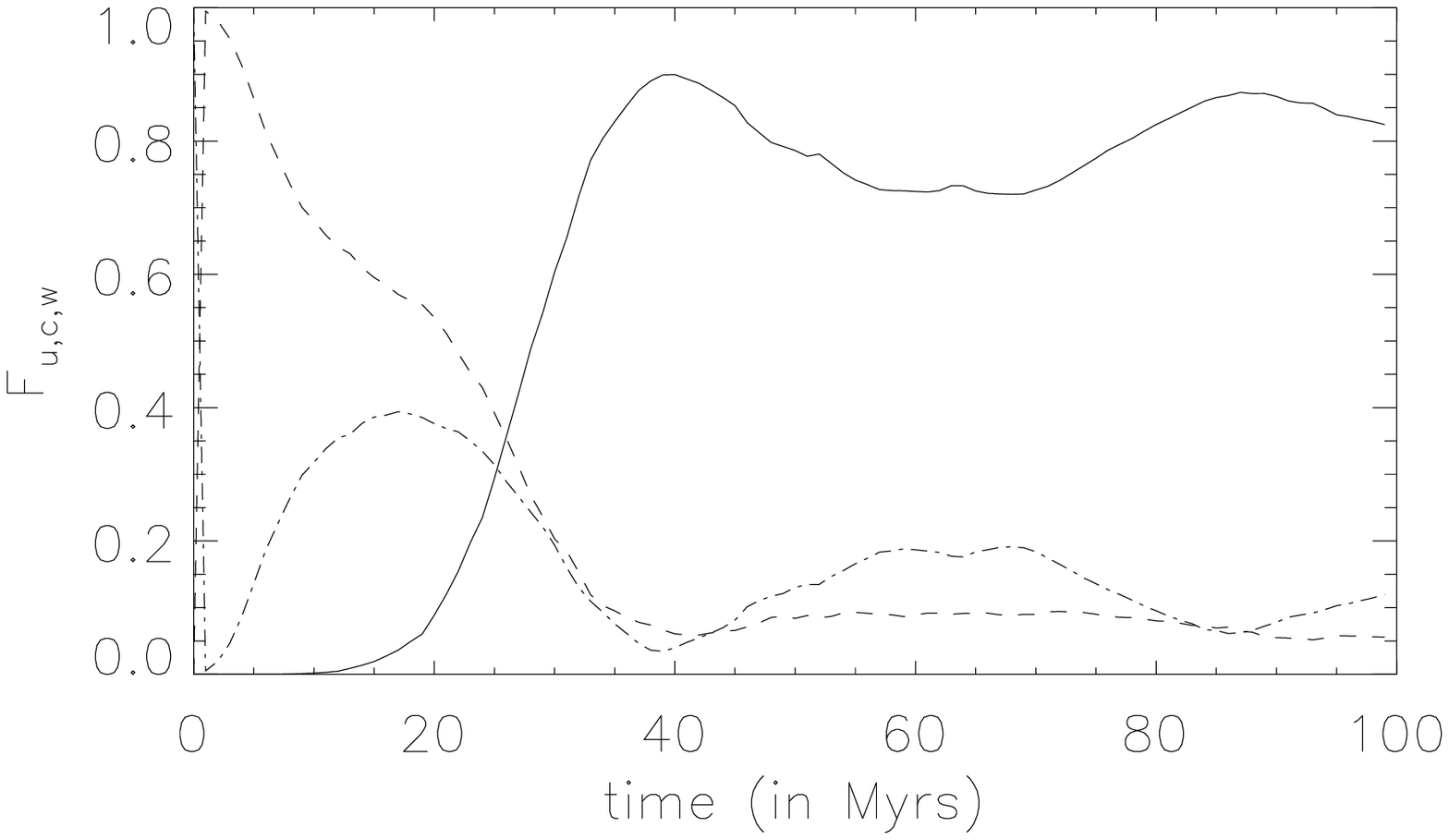} 
\caption{Time evolution of the volume filling factor of the unstable gas (398 K $\lesssim$ $T$ $\lesssim$ 1000 K, dashed line), cold gas ($T < 398$ K, solid line) and warm gas ($T > 10000$ K, dot-dashed line) for the simulation with the set of parameters $(\eta/\eta_{G},\epsilon)=(0.01,0.25)$.}
\label{fig15}  
\end{figure}

\begin{figure}
\plotone{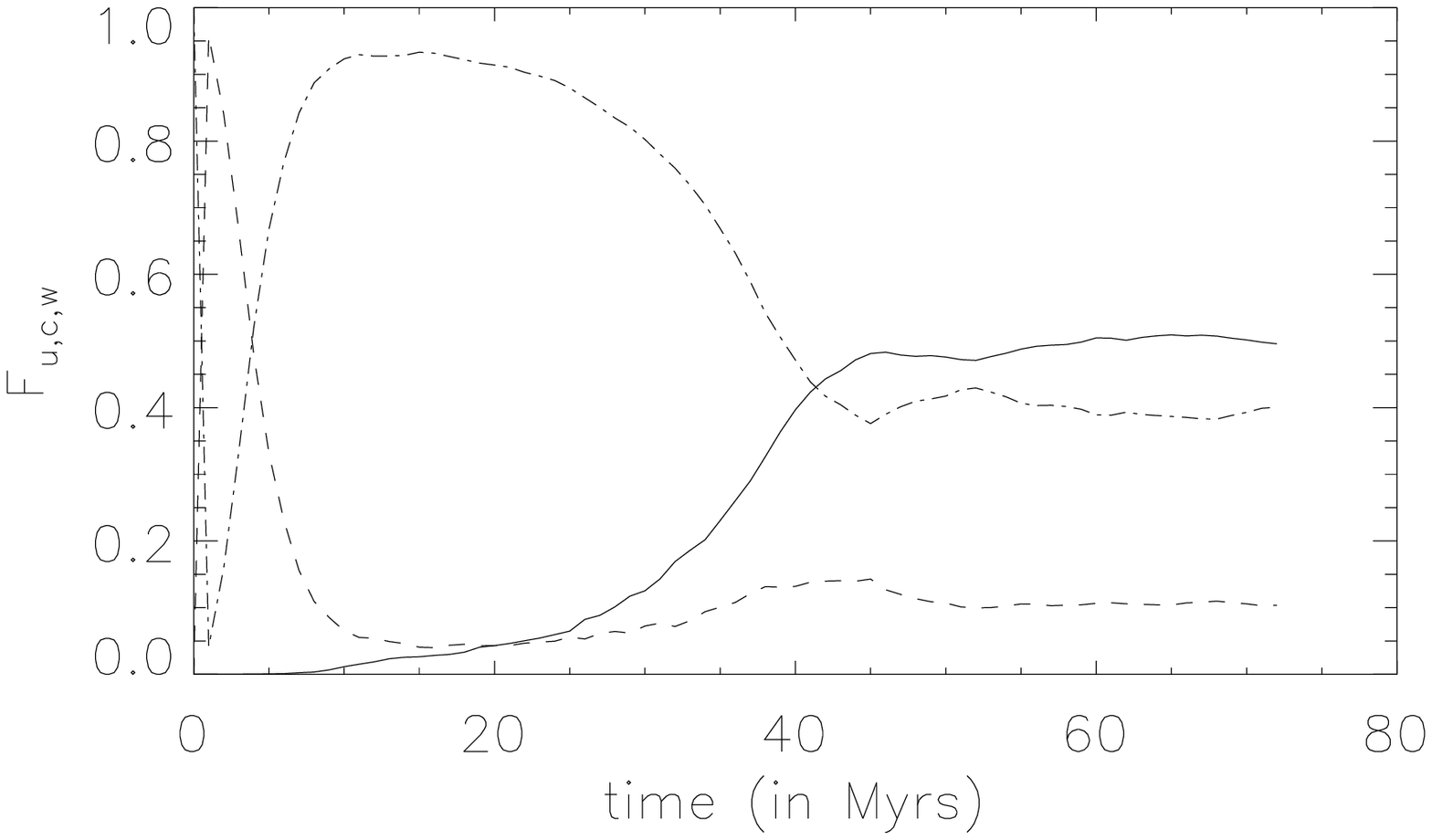} 
\caption{Time evolution of the volume filling factor of the unstable gas (398 K $\lesssim$ $T$ $\lesssim$ 1000 K, dashed line), cold gas ($T < 398$ K, solid line) and warm gas ($T > 10000$ K, dot-dashed line) in the simulation with the set of parameters $(\eta/\eta_{G},\epsilon)=(0.1,0.25)$.}
\label{fig16}  
\end{figure}

\begin{figure}
\plotone{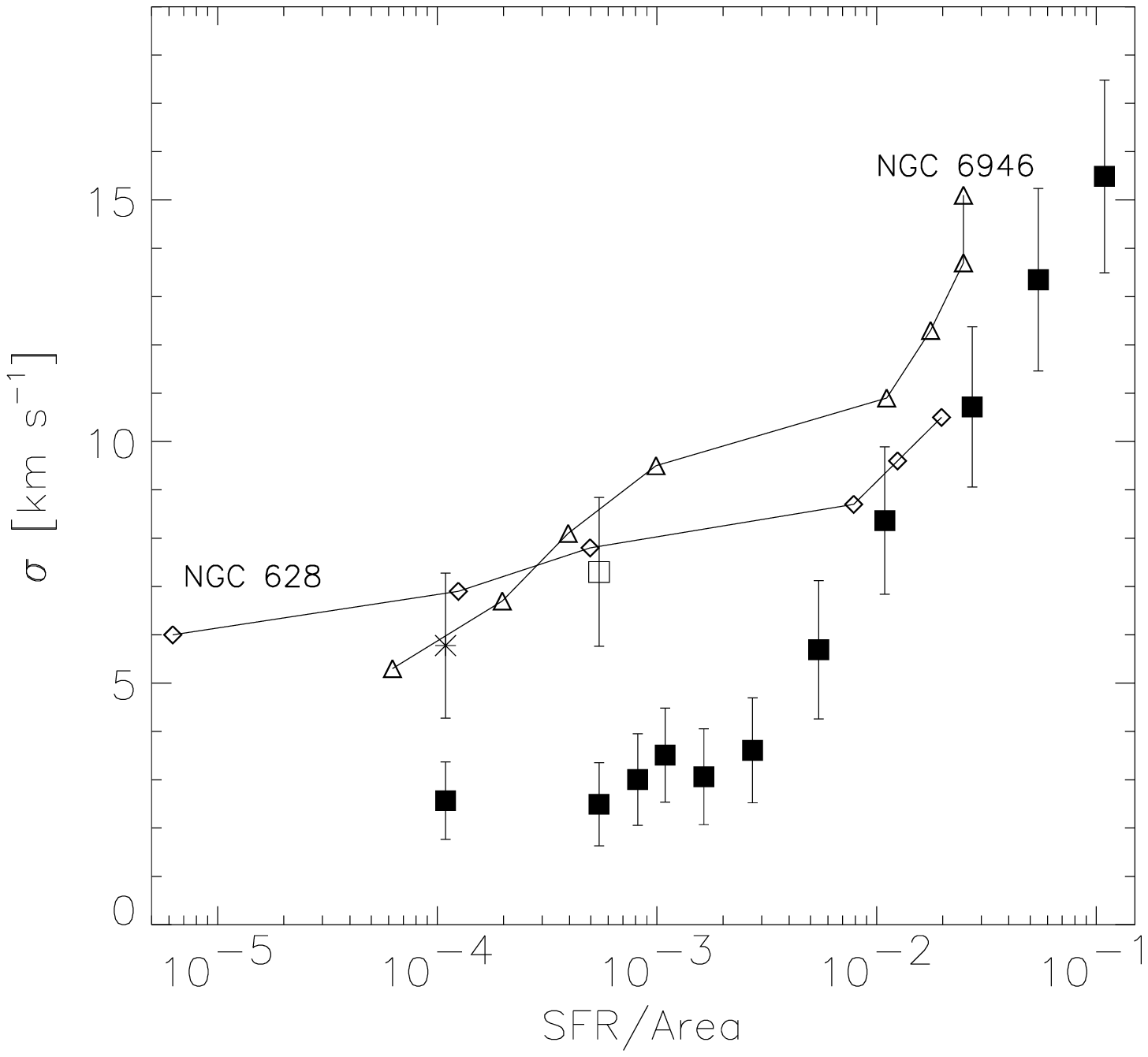} 
\caption{Velocity dispersion of the gas $\sigma$ as a function of the star formation rate per unit area (M$_{\odot}$~yr$^{-1}$~kpc$^{-2}$) (filled squares). open square and star correspond to simulations where the density has been reduced by a factor of 5 and 10, respectively. The displayed data has been obtained using a spectral bin size of 1 km s$^{-1}$. For NGC 628 (open diamonds) and NGC 6946 (open triangles), the velocity dispersion is derived from \ion{H}{i} 21 cm line observations and the star formation rates from H$\alpha$ observations.} 
\label{fig17}  
\end{figure}

\begin{figure}
\plotone{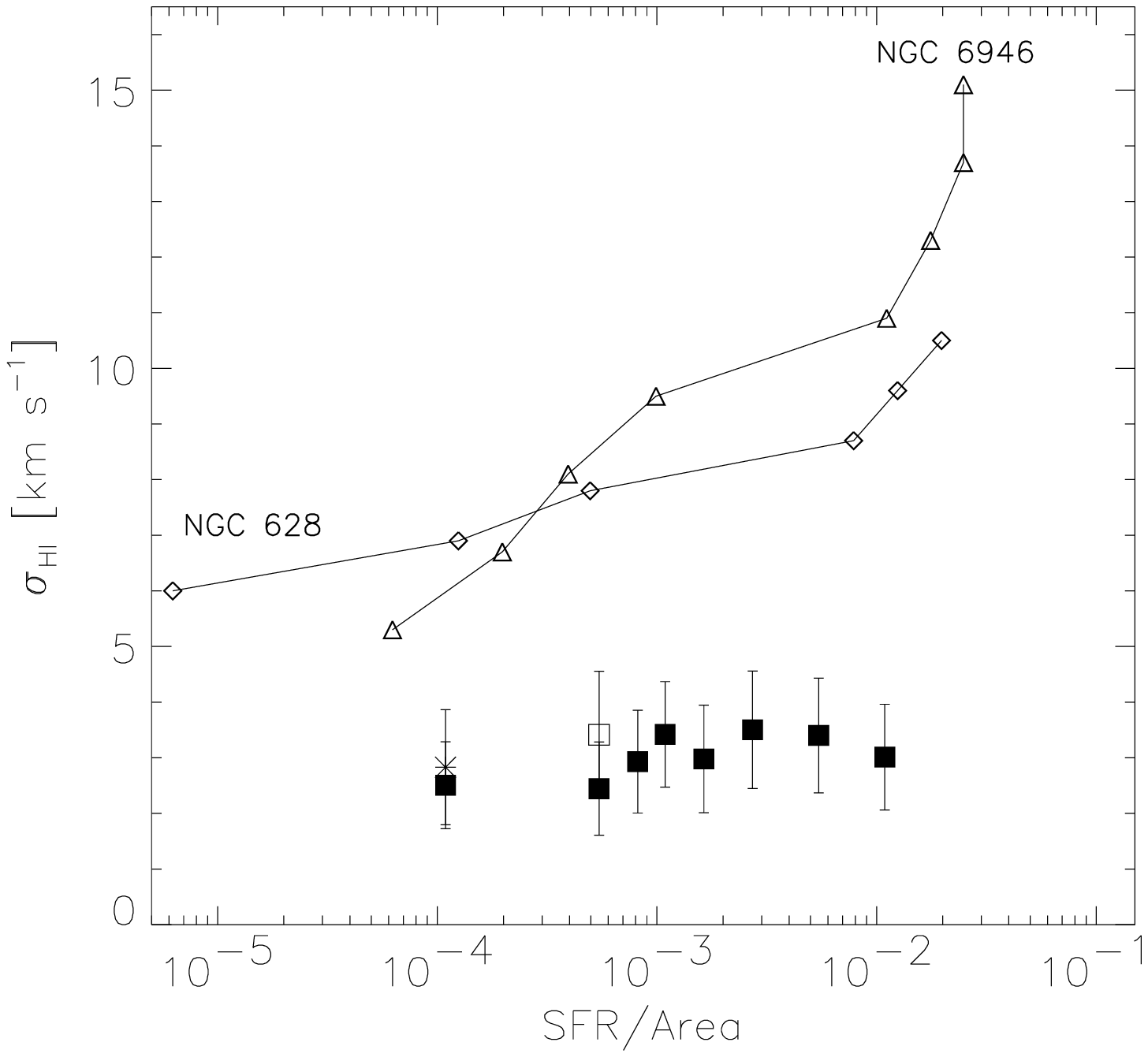} 
\caption{Velocity dispersion of the \ion{H}{i} gas $\sigma_{\ion{H}{i}}$ as a function of the star formation rate per unit area (M$_{\odot}$ yr$^{-1}$~kpc$^{-2}$) (filled squares). open square and star correspond to simulations where the density has been reduced by a factor of 5 and 10, respectively. The displayed data has been obtained using a spectral bin size of 1 km s$^{-1}$. For NGC 628 (open diamonds) and NGC 6946 (open triangles), the velocity dispersion is derived from \ion{H}{i} 21 cm line observations and the star formation rates from H$\alpha$ observations.}
\label{fig18}  
\end{figure}

\begin{figure}
\epsscale{0.8}
\plotone{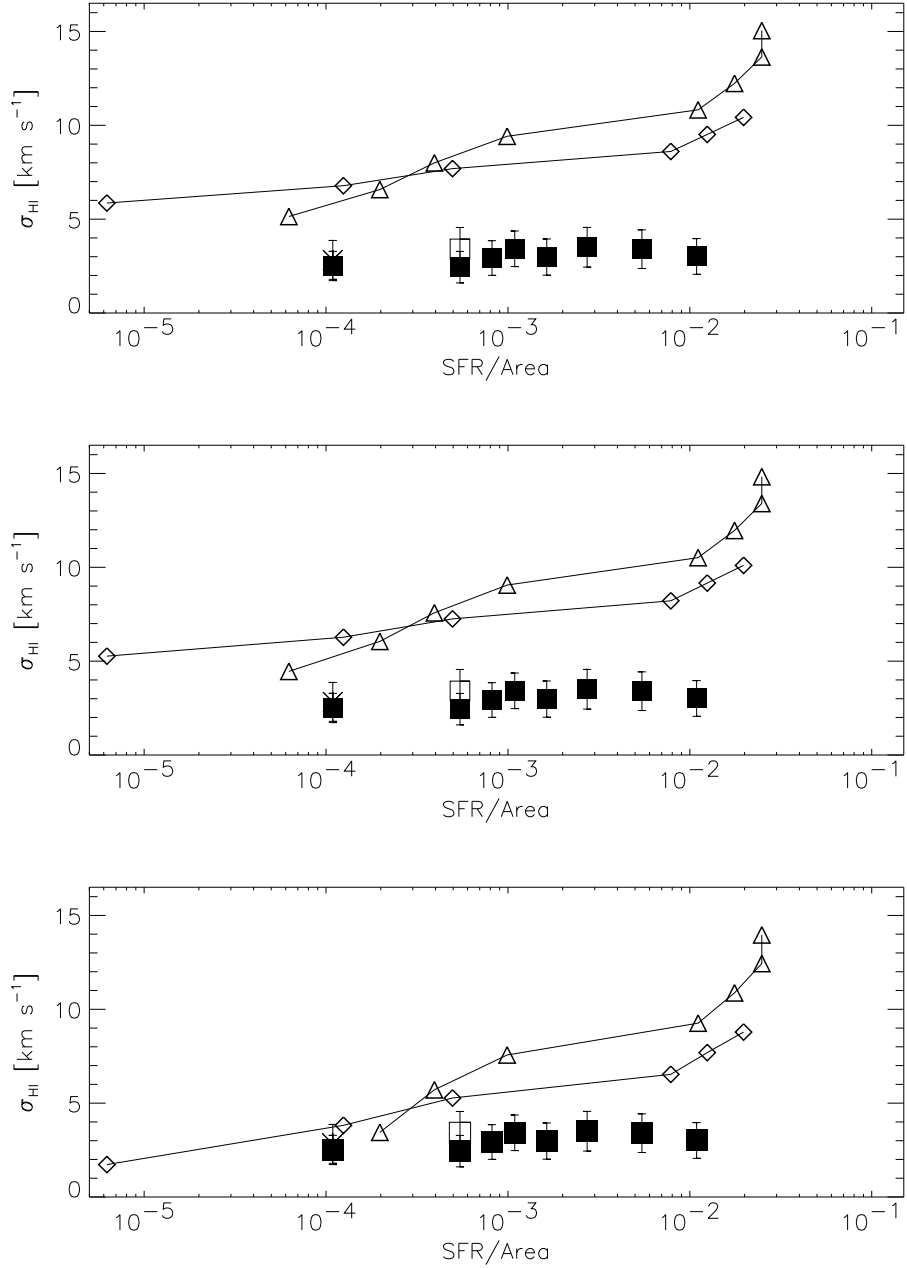} 
\caption{Same as in Fig.~\ref{fig18} but where thermal broadening has been subtracted from the observed velocity dispersion assuming that the \ion{H}{i} gas is at 100 K (top), 500 K (middle) and 2000 K (bottom). Spectral bin size used is 1 km s$^{-1}$.} 
\label{fig19} 
\end{figure}

\begin{figure}
\plotone{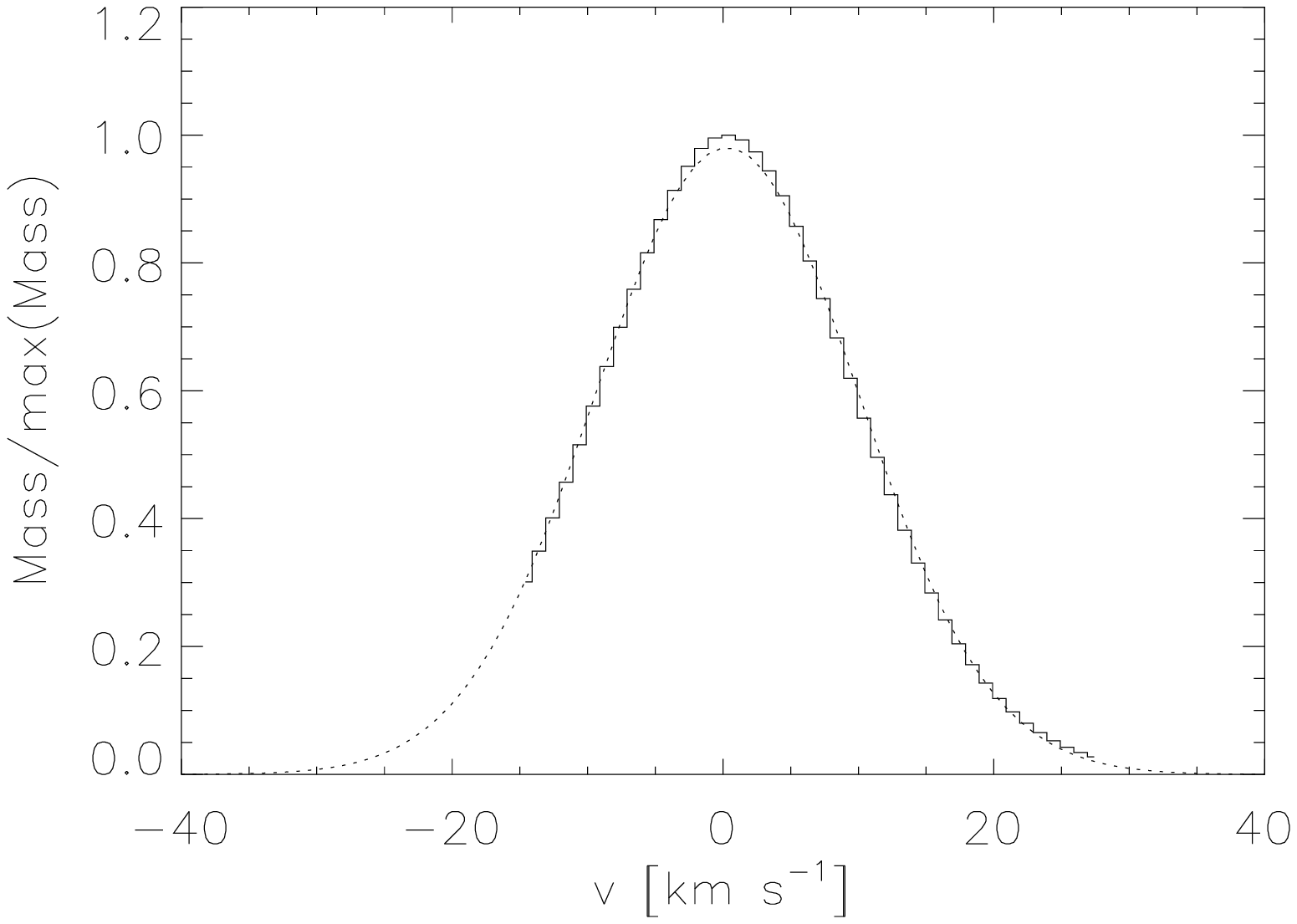} 
\caption{Mass weighted velocity profile for the model with ($\eta/\eta_{G},\epsilon$)=(1,0.25) and corrected for thermal broadening. The spectral bin size of 1 km s$^{-1}$. The profile samples only gas elements with $T \leq 12000$ K and $n \geq 0.25$ cm$^{-3}$. Over-plotted is a Gaussian fit (dotted line).} 
\label{fig20}  
\end{figure}

\begin{figure}
\plotone{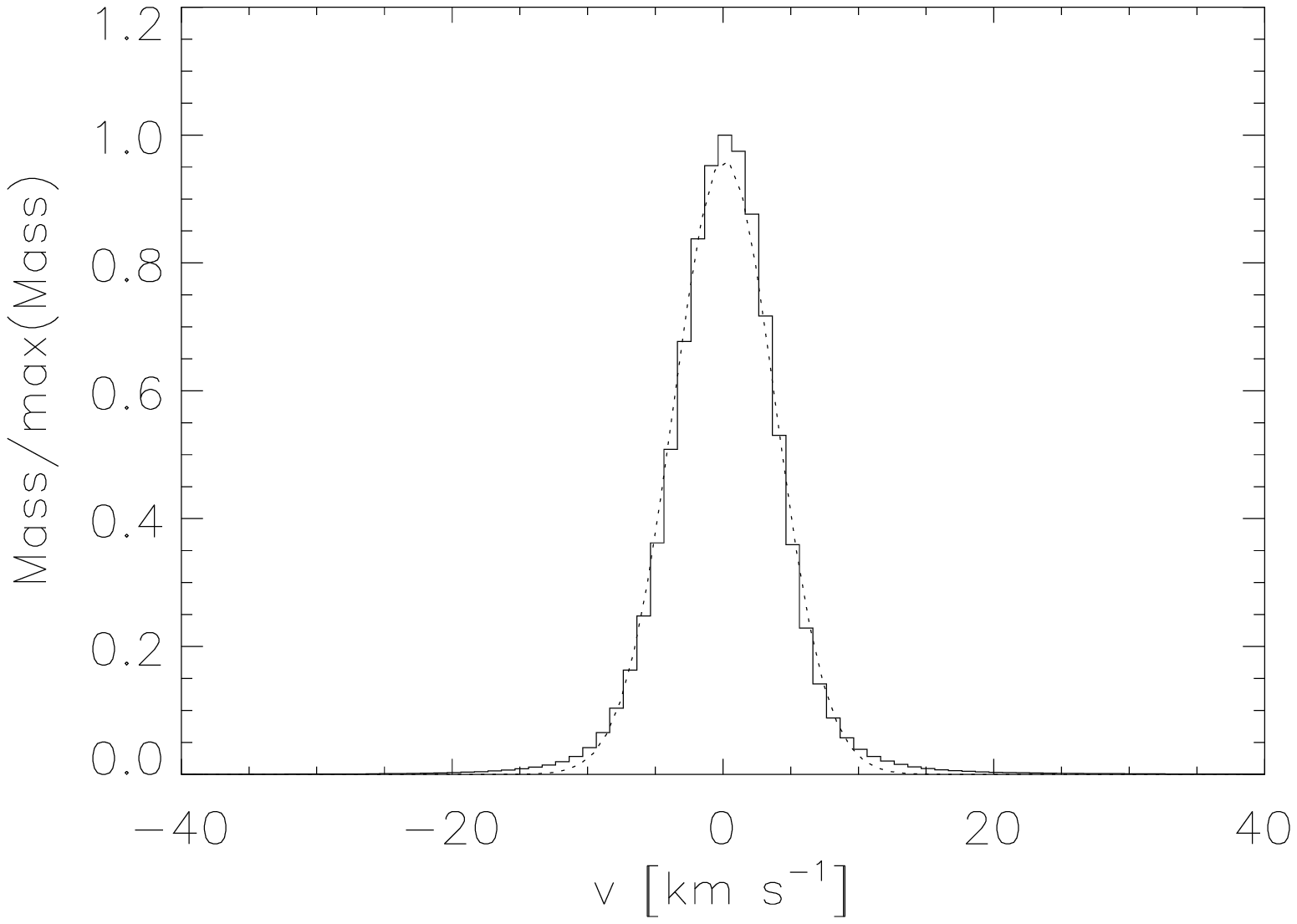} 
\caption{Mass weighted velocity profile for the model with ($\eta/\eta_{G},\epsilon$)=(0.1,0.25) and corrected for thermal broadening. The spectral bin size of 1 km s$^{-1}$. The profile samples only gas elements with $T \leq 12000$ K and $n \geq 0.25$ cm$^{-3}$. Over-plotted is a Gaussian fit (dotted line).} 
\label{fig21} 
\end{figure}

\begin{figure}
\plotone{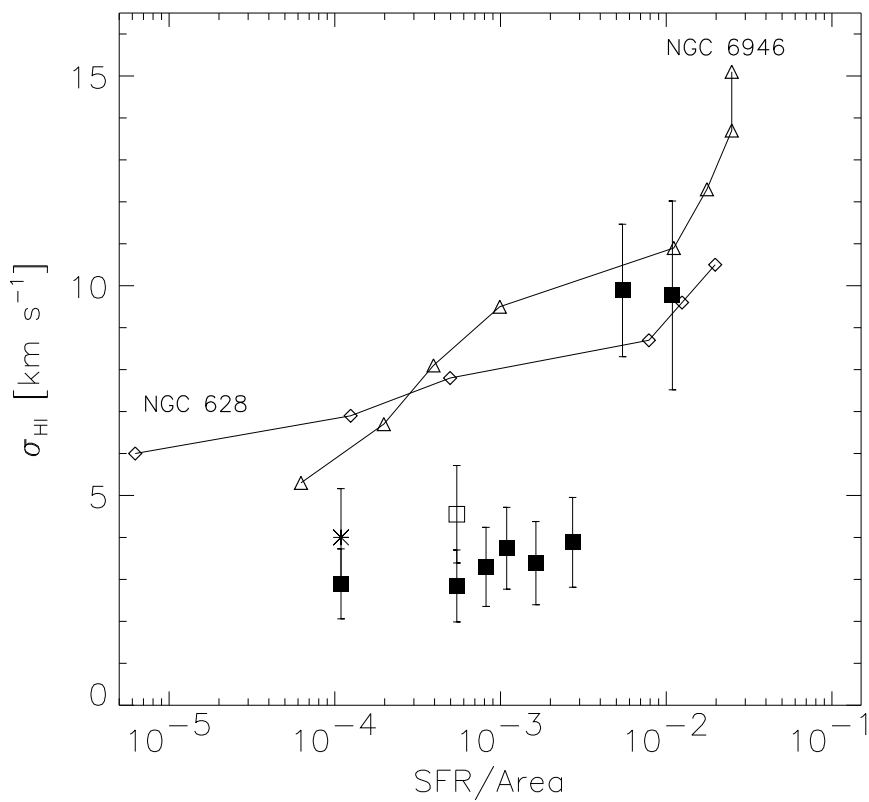} 
\caption{Velocity dispersion of the \ion{H}{i} gas $\sigma_{\ion{H}{i}}$ as a function of the star formation rate per unit area (M$_{\odot}$ yr$^{-1}$ kpc$^{-2}$) (filled squares). The values have been corrected for the effect of thermal broadening. Open square and star correspond to simulations where the density has been reduced by a factor of 5 and 10, respectively. The displayed data has been obtained using a spectral bin size of 1 km s$^{-1}$.} 
\label{fig22}  
\end{figure}

We compare  $\sigma_{\ion{H}{i}}$ to the observations as a function of the SN rate (i.e., SFR rates). $\sigma$ and $\sigma_{\ion{H}{i}}$ have nearly similar values at the low SN rates as most of the gas in this regime is \ion{H}{i} gas. \ion{H}{i} gas is not present in the simulations with $\eta/\eta_{G} \gtrsim 1$. $\sigma_{\ion{H}{i}}$ is observed to be nearly independent of the SN rate, leveling off at $\sim 3$ km$^{-1}$. This value is a factor $2-3$ lower that the velocity dispersion plateau observed in galaxies such as NGC 1058, NGC 628 and NGC 6964 (Fig.~\ref{fig1} and Fig.~\ref{fig18}). However, the set of simulations with $\eta/\eta_{G}=0.1$ and the different feedback efficiencies (i.e, Fig.~\ref{fig11}) suggests that \ion{H}{i} gas velocity dispersions of the order of 5-6 km s$^{-1}$ can be obtained for feedback efficiencies $\epsilon \gtrsim 0.5$, in closer agreement with the observations. For the simulations with our fiducial value of $\epsilon$, we have corrected the \ion{H}{i} line velocity profiles by accounting for the effect of thermal broadening. A reasonable agreement is found between the models and the observations for values of the SFR/Area in the range $5\times 10^{-3}-10^{-2}$ M$_{\odot}$ yr$^{-1}$ kpc$^{-2}$. For smaller values of the SFR, the fact that the velocity dispersions are a factor $\sim 2$ smaller than the observed values could result from the fact that we have underestimated the SN feedback efficiency. Otherwise, it might be an indication of the existence of secondary heating and/or driving mechanisms in the outer parts of galaxies where the star formation rate is low. An investigation of the effects of SN driving in the presence of other physical processes and gas instabilities is essential and is left to future work.  

\acknowledgements

It is a pleasure to thank  Enrique V\'{a}zquez-Semadeni, Fabian Walter, Hans-Walter Rix, Axel Brandenburg, Maarit Korpi, Joop Schaye and Robert Piontek for useful comments and discussion. We are also very grateful to the anonymous referee for interesting comments and suggestions. S. D. would like to thank Hans-Walter Rix and Thomas Henning for the financial support at the MPIA. Calculations have been performed on the MPIA's SGI Origin 2000 located at the Rechnenzentrum of the Max-Planck Gesellschaft, Garching. ZEUS-3D was used by courtesy of the Laboratory of Computational Astrophysics at the NCSA. This research has made use of NASA's Astrophysics Data System Bibliographic Services. 

{}

\clearpage 

\clearpage

\clearpage

\clearpage

\end{document}